\documentclass[sigconf]{acmart}
\usepackage{booktabs} 
\usepackage{tcolorbox}
\usepackage{float}
\usepackage{wrapfig}

\usepackage[font=small]{caption}  
\usepackage{enumitem}
\usepackage{titlesec}
\usepackage{subcaption}
\usepackage{graphicx}
\usepackage[font=small,labelfont=bf]{caption} 
\definecolor{deepgreen}{rgb}{0.0, 0.5, 0.0} 

\AtBeginDocument{%
  }

\setcopyright{acmlicensed}
\copyrightyear{2025}
\acmConference[Co-DESIGN'25]{25}{Nov 16--21, 2025}{}
\acmISBN{978-1-4503-XXXX-X/2025/11}

\begin{document}


\author{Jesmin Jahan Tithi}
\affiliation{%
  \institution{Intel}
  \city{Santa Clara}
  \state{California}
  \country{USA}
}
\email{jesmin.jahan.tithi@intel.com}

\author{Hanjiang Wu}
\affiliation{%
  \institution{Georgia Institute of Technology}
  \city{Atlanta}
  \state{Georgia}
  \country{USA}
}
\email{hwu419@gatech.edu}

\author{Avishaii Abuhatzera}
\affiliation{%
  \institution{Intel}
  \city{Haifa}
  \country{Israel}}
\email{avishaii.abuhatzera@intel.com}

\author{Fabrizio Petrini}
\affiliation{%
  \institution{Intel}
  \city{Santa Clara}
  \state{California}
  \country{USA}
}
\email{fabrizio.petrini@intel.com}

\renewcommand{\shortauthors}{jesmin jahan tithi et al.}

\title{Scaling Intelligence: Designing Data Centers for Next-Gen Language Models}

\begin{abstract}

The explosive growth of Large Language Models (LLMs)—such as GPT-4 with 1.8 trillion parameters—demands a fundamental rethinking of data center architecture to ensure scalability, efficiency, and cost-effectiveness. Our work provides a comprehensive co-design framework that jointly explores FLOPS, HBM bandwidth and capacity, multiple network topologies (two-tier vs. FullFlat optical), the size of scale-out domain, and popular parallelism/optimization strategies used in LLMs. We introduce and evaluate Fullflat network architectures, which provide uniform high-bandwidth, low-latency connectivity between all nodes, and demonstrate their transformative impact on performance and scalability. Through detailed sensitivity analyses, we quantify the benefits of overlapping compute and communication, leveraging hardware-accelerated collectives, widening the scale-out domain, and larger memory capacity. Our study spans both sparse (mixture of experts) and dense transformer-based LLMs, revealing how system design choices affect Model FLOPS Utilization (MFU = Model flops per token \(\times\) Observed tokens per sec/Peak flops of the hardware) and overall throughput. For the co-design study, we utilized an analytical performance modeling tool capable of predicting LLM runtime within 10\% of real-world measurements. Our findings offer actionable insights and a practical roadmap for designing AI data centers that can efficiently support trillion-parameter models, reduce optimization complexity, and sustain the rapid evolution of AI capabilities.

\end{abstract}

\keywords{Co-design, Data center, GPT3, GPT-4, LLM, MOE, Fullflat Optics, Fullflat Network, Performance Analysis, Co-packaged optics}

\settopmatter{printacmref=false}
\setcopyright{none}
\renewcommand\footnotetextcopyrightpermission{}
\pagestyle{plain}
\maketitle

\section{Introduction}

Future data centers must be equipped to handle the demands of multi-trillion parameter large language models (LLMs) like OpenAI's GPT-4 with mixture of experts transformer models \cite{cai2024survey} featuring 1.8 trillion parameters. These models require advanced data center designs with substantial compute power, high-bandwidth memory (HBM), and scalable, low-latency networks. The cost of training GPT-4 far exceeds that of GPT-3, underscoring the need for efficient infrastructure. The widespread adoption of LLMs across industries pushes data center co-design priorities, challenging throughput, latency, and reliability. Current systems often achieve less than 50\% Model FLOPS Utilization (MFU)\footnote{takes into account the required operations for the forward and backward passes through the model and does not include recomputation in the measured FLOPs.} \cite{ultrascale_playbook,shoeybi2019megatron}, highlighting the need for optimization.

Our co-design analysis focuses on data center architectures needed to efficiently run large-scale models like GPT-4 due to its significant impact on the infrastructure and operational demands of modern data centers. Training GPT-4 required approximately 25,000 Nvidia A100 GPUs over 90-100 days, using a dataset of 13 trillion tokens. Holding model parameters, gradients, and optimizers demands around 461 A100 GPUs, each with 80 GB of HBM, assuming 20 bytes per parameter for BF16 compute and FP32 grad accumulation \cite{ultrascale_playbook}.

Today's AI data centers are built with clusters of GPUs that are organized into nodes with a limited number of GPUs (e.g., 8) inside each node. These GPUs are interconnected through high-speed links such as NVLink (Nvidia), XeLink (Intel) \cite{jiang2022intel}, or Infinity Fabric (AMD) \cite{amd2024mi300x}. Additionally, the nodes are connected via lower-bandwidth networks like Ethernet or InfiniBand throughout the data center. This setup can be viewed as a two-tier network consisting of high-bandwidth and low-bandwidth domains (HBD and LBD for short). The difference in throughput between these two is significant; for example, NVLink provides 450 GB/s, whereas Ethernet offers 50 GB/s. HBD is typically utilized for communication-intensive tensor parallelism, while LBD facilitates data and pipeline parallelism.

The upcoming Nvidia Grace Blackwell GPU (i.e., GB200, GB300) enhances the HBD by expanding its capacity from 8 to 72 GPUs with an NVLink bandwidth of 900 GB/s, doubling previous capabilities. The Nvidia Rubin series plans further expansion to 144 or 576 GPUs with an NVLink bandwidth of 3600 GB/s by 2027 \cite{Patel2025NVIDIA}. We use ``GPU" and ``node" interchangeably, with ``GPU" potentially replaced by TPU, CPU, XPU, or other accelerators.

Optimizing data centers for large-scale models like GPT-4 is a complex and challenging task due to the numerous variables that need to be chosen and fine-tuned, from both system architecture and software algorithm perspectives. This creates a multifaceted optimization landscape. From a system architecture perspective, key factors include compute nodes, storage, network infrastructure, bandwidth, and latency. Support for high-performance collective libraries, such as Nvidia's SHARP \cite{nvidia_sharp,graham2020scalable}, is important for communication and synchronization.

Within each node, critical parameters that directly influence LLM performance include the floating-point operations per second (FLOPS) capabilities of the GPUs across various precisions (FP64, FP32, FP16, FP8, FP4, etc.) and their efficiency at different operational sizes. The capacity, bandwidth, and transfer efficiency of the GPU's high-bandwidth memory (HBM or Tier 1 memory) are also important metrics. Furthermore, the capacity, bandwidth, and transfer efficiency of the attached CPU's DDR memory (or Tier 2 memory), which is often used for offloading inactive model components and checkpointing, and the bandwidth of data transfer between these memories and the GPU, are critical factors to consider \cite{ultrascale_playbook}.

On the software algorithm side, a multitude of optimization parameters require careful tuning. These include various forms of parallelism, such as data parallelism (DP), pipeline parallelism (PP), sequence parallelism (SP), context parallelism (CP), tensor parallelism (TP), expert parallelism (EP), and expert shading (ES). Additional considerations involve determining the optimal batch size and micro-batch size, the degree of pipeline interleaving, and other strategies to reduce pipeline bubbles and inefficiencies. Decisions on quantization and data types for computation, and whether to employ kernel fusion by combining activation computation with preceding computations have a performance impact. The choice of attention, such as multi-head (MHA), Multi-head Latent (MLA), multi-query (MQA), or group query (GQA) — also plays a role, as does the decision on activation recomputation (none, full, or attention-only) in determining the performance of LLM while running in a data center cluster. Deciding to do full recompute by default adds 30\% overhead in the total time, where it saves significant storage needs by avoiding storing activations, thus, allowing it to run on fewer nodes. Some of the recomputation can be avoided by offloading inactive layer's weights, activations, and optimizers to CPU/Tier 2 memory and transferring them back when needed to the GPU memory, and the decision on whether to do this or not can have an impact on the performance and storage need \cite{ultrascale_playbook}.

Additional considerations include optimizer sharding techniques such as ZERO-1, which partitions the optimizer, and ZERO-2, which extends this to include both optimizer and gradients. FSDP/ZERO-3 further expands on this by sharding the optimizer, gradients, and weights. Overlapping communication with computation in tensor parallelism (TP) and data parallelism (DP) is also important. These optimizations have specific communication requirements, and the network's type and characteristics can affect perceived performance and scalability. The choice of collective operations for synchronizing and exchanging values across GPUs also impacts performance; for example, the all-reduce operation can be implemented as a combination of reduce-scatter and all-gather, which can be divided into chunks and interleaved with computation. When executed effectively, this interleaving can mask the data transfer cost, though it typically requires high-performance interconnects to achieve optimal efficiency. Collectives can be implemented in hardware with smart switches or in software. The type of collective operation and usage pattern influence LLM performance in a data center and should be considered during both model implementation and data center design \cite{ultrascale_playbook,nvidia_sharp,graham2020scalable}.

\begin{figure}
    \centering
    \includegraphics[width=0.4\textwidth]{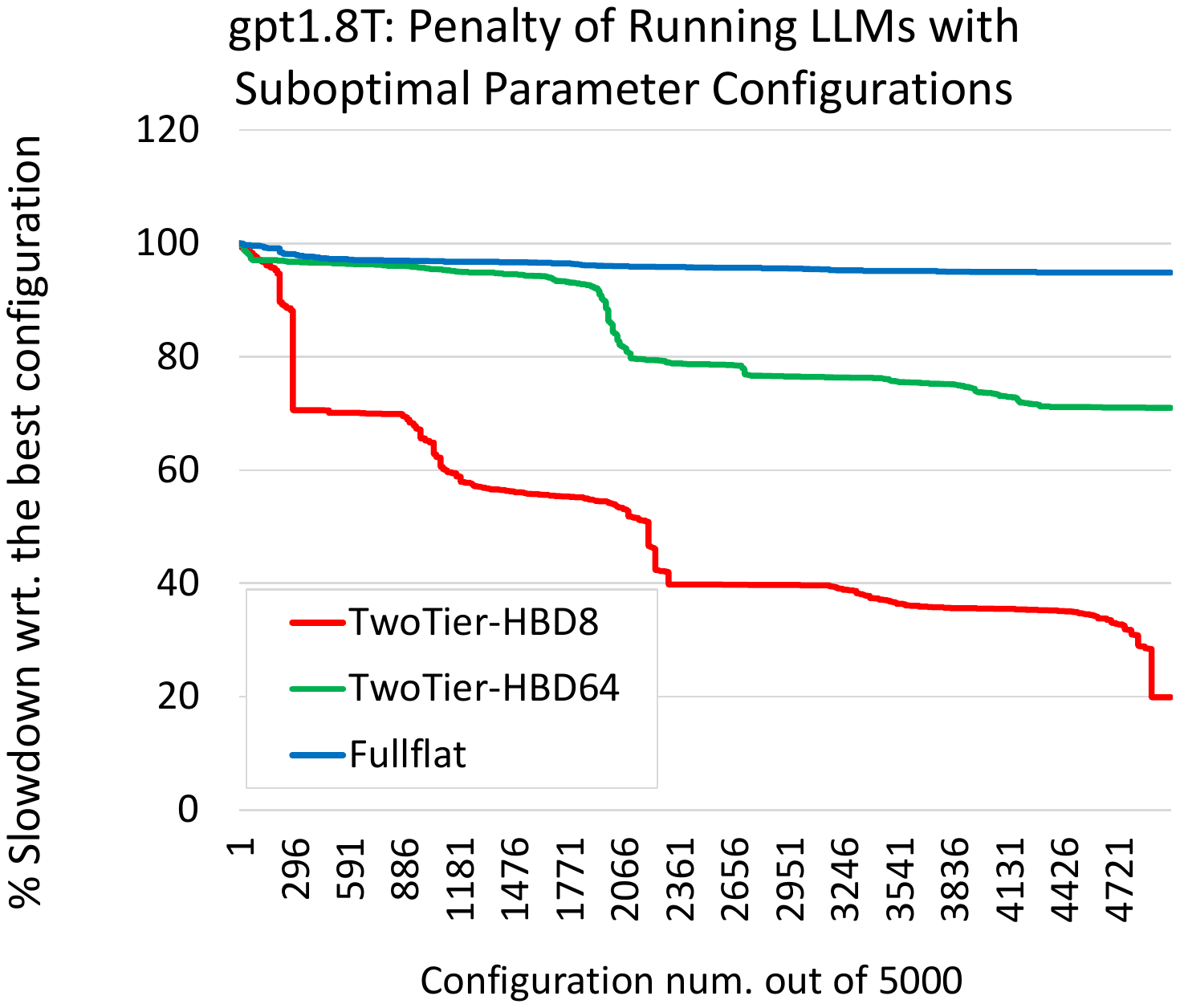} 
    \caption{Running LLMs with suboptimal parameter configurations on different system configurations. Here: TwoTier-HBD8 is today's system, TwoTier-HBD64 are systems that will be available soon (within a year) and an all-to-all FullFlat system with flat bandwidth (same interconnect) for scale-up (HBD) and scale-out (LBD) domains. System size: 65,535 GPUs.}
    \label{fig:opt}
\end{figure}

The choice of optimization parameters significantly impacts performance, and their interactions are complex \cite{ultrascale_playbook}. Accurately estimating the performance of a GPT model requires a sophisticated tool that models the LLM's components and comprehensively assesses the impact of each parameter. Such a tool can perform exhaustive searches to identify the optimal parameter combinations for a given system architecture, maximizing performance. Running LLMs with suboptimal configurations can lead to substantial performance penalties, with performance reductions of up to 80\% (see Figure \ref{fig:opt}) among the top 5,000 configurations, using the same number of GPUs. Apparently, performance can vary widely based on the system setup and optimization choices. This highlights the importance and complexity of the optimization process.

In this paper, we utilize a Python-based detailed analytical tool to model the performance of some popular LLMs. This tool incorporates various parameters to predict the runtime of LLM training on a given system architecture. It builds upon the open-source framework \cite{isaev2023calculon}, enhancing its capabilities to include components essential for Mixture of Experts (MOE) \cite{sanseviero2023moe}, expert parallelism, and other features relevant to GPT-4 style LLMs. Our tool has been demonstrated to predict runtimes within 10\% of actual system performance. Since this paper focuses on co-design insights and lessons learned from evaluating potential future data center architectures, the full technical details of the tool and its validations will be presented in a separate paper.

{\bf{Contributions:}}
Our work provides a comprehensive co-design framework that jointly explores compute, memory, networking, and storage capabilities needed for efficient training of next-generation LLM models. We explore traditional two-tiered and futuristic fully optical Fullflat networks. Nvidia H100, GB200, and Rubin systems represent two-tiered systems; thus, our work is directly applicable to next-gen architectures. By exhaustively evaluating thousands of hardware-software configurations, we reveal non-obvious trade-offs—such as diminishing returns from enlarged High Bandwidth Domain (HBD) once MoE communication fits in the HBD, and how FullFlat enables tensor parallelism (TP) beyond a node boundary, contrary to prevailing assumptions in current real systems. We share insights on required future data center capabilities regarding the training of large-scale LLMs.

\begin{itemize}[leftmargin=7pt] 
\item {\bf {Extended Analytical Framework for MoE LLMs}}:We substantially extend the open-source Calculon simulator to support large-scale MoE models, dynamic all-to-all routing, SwiGLU, and relevant communication operations—capabilities absent from existing tools. These enhancements enable accurate, early-stage exploration for next-generation LLMs, combining key parallelization and optimization features from Megatron, DeepSpeed , and NEMO. Our framework quantifies how modern MoE architectures (with expert-parallel MLPs and dynamic routing) alter performance scaling, guiding both system architects and practitioners before infrastructure commitments.

\item {\bf {FullFlat Topology: A New Optimization Frontier}}: We present the first systematic evaluation of FullFlat optical topologies for LLM workloads. Its high-radix, all-optical fabric delivers full bisection bandwidth, adaptive routing, and topology-aware GPU placement, enabling novel parallelization and communication-overlap strategies beyond what two-tier networks support. FullFlat improves MFU for both sparse and dense models, showing the highest utilization among all evaluated systems. It also reduces sensitivity to software optimizations, easing performance tuning and mitigating``performance islands." While GPU memory and model-specific constraints may limit certain optimizations, FullFlat cuts switch count, lowers multi-hop overhead, and can achieve 20–30\% TCO savings, all while enhancing reliability and serviceability.

\item {\bf {Network Scalability Challenges for Multi-Trillion Parameter GPT Models}}: Our co-design analysis demonstrates that, despite the availability of high-end accelerators (i.e., GPUs, TPUs), and CPUs with multi-petaflop computational capabilities, network scalability emerges as a critical bottleneck when training multi-trillion parameter GPT models across extensive compute arrays. Our findings underscore the advantages inherent in a unique high radix, low diameter (e.g., 2D HyperX \cite{domke2019hyperx}, PolarStar \cite{lakhotia2024polarstar}), high-speed co-packaged optics enabled Fullflat network, with the same bandwidth connection between any two nodes where the entire network could practically operate as an HBD.  

\item {\bf {Impact of HBD on Optimization Choices}}: We investigate the optimal balance between HBD and LBD across various GPT architectures, assessing the adequacy of HBD. We analyze how different parallelization strategies—TP, DP, PP, EP, ES—are influenced by the HBD/LBD design and their impact on overall model performance.

\item {\bf {Performance Sensitivity with Network, Compute, and Memory}}: We analyze performance sensitivity with different system parameters and summarize findings. Sensitivity depends on the application's arithmetic, memory, and network intensities \cite{Ridgeline}. We summarize the impact factor of different resources for the GPT models studied in this paper. Such impact factor can help one to decide with resources should be priorited based on return on investment (ROI) metric. 

\item {\bf {Collective Communication and Overlap Strategies}}: We evaluate the performance benefits of hardware-accelerated collective libraries (e.g., NVIDIA SHARP) and quantify the impact of overlapping communication with computation. We also contrast sparse and dense models to show how these strategies vary by workload.

\item {\bf {GPU Memory Capacity as a Critical Bottleneck}}: We emphasize the importance of GPU HBM capacity in reducing memory pressure, lowering GPU count for model training, and improving performance—especially for dense models like GPT-3 that lack the sparsity advantages of MoE systems.

\item {\bf{Comparison of Sparse and Dense Models}}: MoE pre-training with dense model distillation is a rising trend. Thus, in co-design it's important to build a system that can serve both. We compare the adaptability and performance of data centers optimized for sparse models when applied to dense models like GPT-3.
\end{itemize}

\section{Prior Work and Background}
Training trillion-parameter LLMs demands significant computational, memory, and networking resources. Key goals in data center design include enhancing performance, improving reliability, ensuring fault tolerance, increasing sustainability, boosting power efficiency, and reducing the total cost of ownership (TCO). A balanced approach is essential for optimizing resource utilization and highest ROI. In this section, we provide background on the co-design setups and tools used in our study. 

\subsection{Prior Work}
Bambhaniya et al. \cite{bambhaniya2024demystifying} introduced GenZ, an analytical tool that clarifies the relationship between LLM {\bf{inference}} performance and various platform design parameters. GenZ models the compute, memory, and interconnect requirements of advanced LLMs like LLaMA and GPT-4 across diverse serving scenarios, projecting the hardware capabilities needed to support future models potentially exceeding hundreds of trillions of parameters. Complementing this, Isaev et al. \cite{isaev2023calculon} developed Calculon, a parameterized analytical performance model that aids in the high-level co-design of systems and transformer-based LLM {\bf{training}}. Calculon, derived from a comprehensive survey of performance optimizations, captures application characteristics, hardware attributes, and implementation strategies to guide algorithm-architecture co-design studies. Together, these tools offer a robust framework for the co-design of next-generation data centers optimized for efficient and scalable LLM deployment. Our work builds upon Calculon, extending it to accommodate Mixture of Experts (MoE) type LLMs and employing it in data center co-design studies. 

A significant focus has been on network design, where Domke et al. \cite{domke2019hyperx} introduced the HyperX topology, providing efficient interconnects for large-scale data centers. Similarly, Lakhotia et al. \cite{lakhotia2024polarstar} proposed the PolarStar topology, which emphasizes low-diameter, high-bandwidth networks to enhance scalability. We explored innovative data center architectures featuring co-packaged optics for both scale-up and scale-out domains \cite{Cerebras2025DARPA,Lightmatter2025PassageL200,Yin2013LIONS,Moore2025CrucialOptical}, achieving Fullflat bandwidth across both domains, unlocking numerous benefits absent in today's two-tiered networks. We discuss the types of data centers required in the near future to effectively serve current and envisioned LLM models.

\subsection{Models and Their Parameters}

\subsubsection{Model Components:}
\begin{figure}[tbph]
    \centering
	{
    \includegraphics[width=0.4\textwidth]{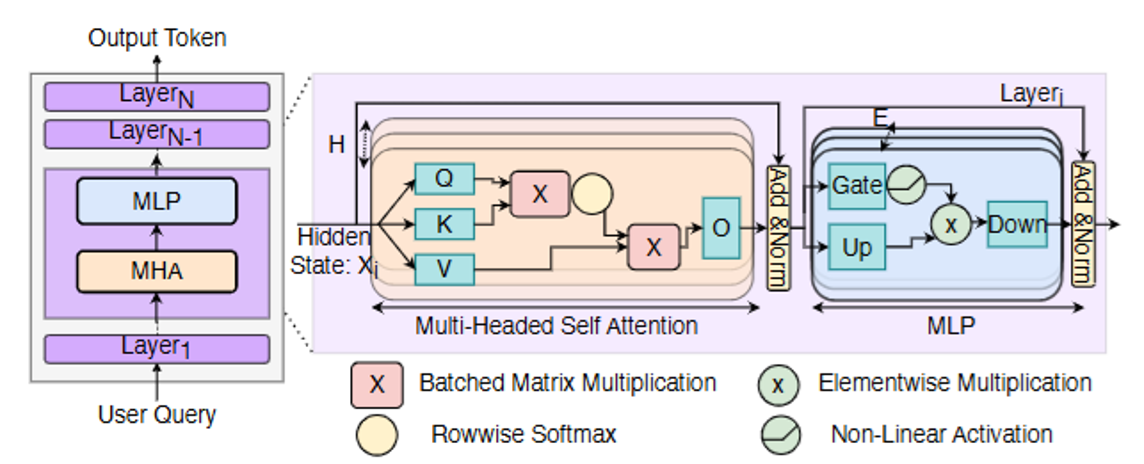}
	
	\includegraphics[width=0.4\textwidth]{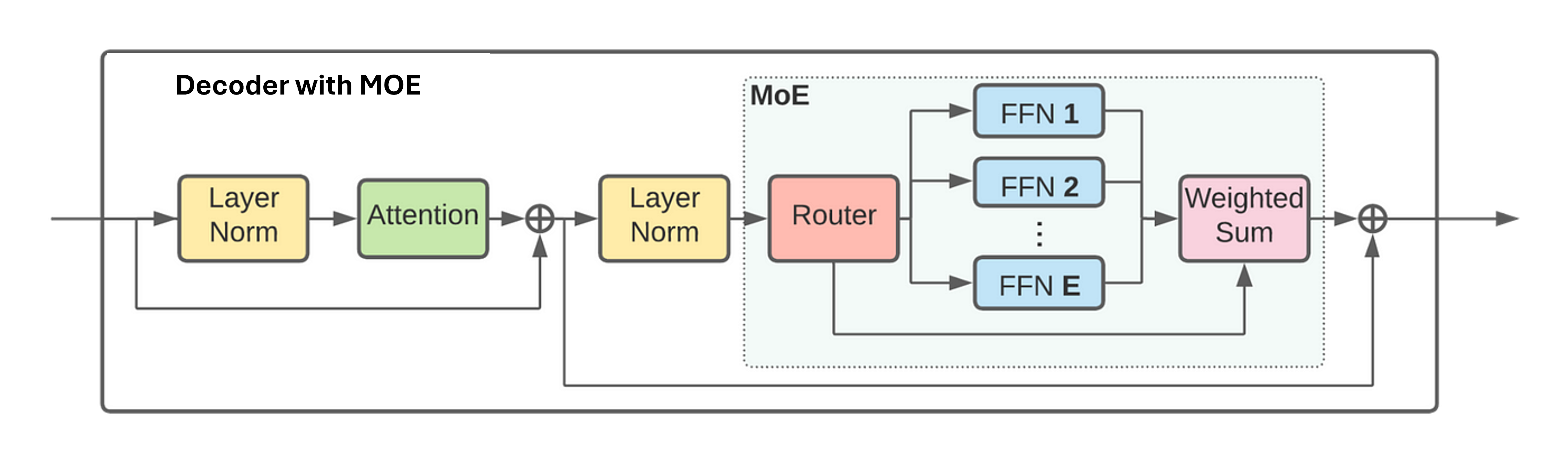} 
	}
    \caption{(Top) Typical LLM components. Adopted from \cite{bambhaniya2024demystifying}. \\(Bottom) GPT-4 Sketch Diagram. }
    \label{fig:LLM}
\end{figure}

GPT models are built by stacking transformer decoder layers, each comprising multi-head self-attention (MHA) and multi-layer perceptron (MLP) blocks. In each layer, the input sequence is transformed into Query (Q), Key (K), and Value (V) activations, divided into Head (H) chunks for parallel computation across the Tensor Parallel (TP) dimension. Attention scores are computed via batch matrix multiplication on Q and K, followed by scaling and softmax, then multiplied with V and projected through a linear layer. The MHA output is added to the initial input and normalized (Figure \ref{fig:LLM} (Top)).

The MLP module includes three linear layers: FFup (feed-forward up projection), FFgate (feed-forward gating), and FFdown (feed-forward down projection). FFup and FFgate project the input from dimension $h$ to $4h$. FFgate's output is activated non-linearly and element-wise multiplied with FFup's output. FFdown then projects this result back to the original dimension. The MLP output is added to its input and normalized, serving as the input to the next decoder layer \cite{bambhaniya2024demystifying}.

A mixture of Experts (MoE) represents a special class of LLMs that incorporate `E’ ``expert" MLP layers, from which `topK’ experts are selected for each input token. GPT models employing MoE (e.g., GPT-4) are referred to as sparse LLMs, as each token is processed by only topK experts instead of all `E’ experts. In contrast, dense language models (e.g., GPT-3) can be seen as a specific instance of MoEs, where `E' = `topK' = 1, indicating a single expert MLP layer utilized for every input token. By leveraging multiple experts and selectively activating a subset for each token, MoEs effectively scale model capacity while maintaining efficient computation and memory usage \cite{ultrascale_playbook,bambhaniya2024demystifying}.

In this research, we study GPT-4 models and contrast them with GPT-3. Although GPT-4's parameters are not officially disclosed, sources \cite{semianalysis2023gpt4, MOE2023gpt4, Reddit, bambhaniya2024demystifying} estimate 1.8 trillion parameters across 120 layers, using a MoE architecture with 16 experts, each having about 111 billion parameters. GPT-4's sparse MLPs process only a subset of tokens per expert. Figure \ref{fig:LLM} (Bottom) illustrates a GPT-4 model, where a trainable router directs tokens to `topK' experts after the self-attention and add-normalize blocks, with results aggregated for the next decoder layer. Prior work Lina \cite{li2023accelerating} identified MoE layer computation and communication as the most time-intensive components of MOE based LLM execution. 

\subsubsection{Optimization Parameters:}
Enhancing LLM training performance involves various optimization techniques, each interacting uniquely with system resources. Our co-design study focuses on optimizations detailed in Table \ref{tab:optimization_techniques}, covering common strategies and their impacts on training performance. Table \ref{tab:optimization_techniques} gives an indication of the complexity of the optimization landscape. Researchers have also explored methods like alternative optimization techniques \cite{jiang2024megascale},\cite{ainslie2023gqa}, \cite{bi2024deepseek}, \cite{dao2022flashattention} and techniques to improve all-to-all communication, as used in DeepSeek \cite{bi2024deepseek}. Future work will analyze these additional techniques.

\begin{table*}[tbph]
\centering
\resizebox{\textwidth}{!}{
\begin{tabular}{|p{19em}|p{65em}|}
    \toprule
    \textbf{Optimization/Method} & \textbf{Description, Impact on Training Performance, and Considerations} \\
    \midrule
    Data Parallelism (DP) & Distributes data across GPUs, each maintaining a copy of the model. Enhances training speed for large datasets but necessitates gradient synchronization. Memory savings are achieved by reducing local batch size.\\
    \midrule
    Pipeline Parallelism (PP) & Allocates model layers across GPUs, allowing simultaneous processing of different pipeline stages. Decreases memory usage per device but may introduce pipeline bubbles. Memory savings are specific to model parameters. \\
    \midrule
    Tensor Parallelism (TP) / Sequence Parallelism (SP) & Splits individual layers across GPUs, facilitates the training of very large models but increases inter-GPU communication. Memory savings apply to model parameters and activations, involving hidden dimensions or sequence length, and require high bandwidth communication. \\
    \midrule
    Expert Parallelization (EP) & Distributes expert sub-networks across GPUs in MoE models, enabling larger models but adding routing communication overhead. \\
    \midrule
    Expert Sharding (ES) & Similar to TP, divides experts across GPUs, reducing memory requirements for large models. \\
    \midrule
    Pipeline Interleaving & Enhances pipeline efficiency by overlapping computations, potentially increasing throughput with similar memory usage.\\
    \midrule
    Batch Size & Total samples processed in one iteration; larger sizes enhance stability and throughput but demand more memory. \\
    \midrule
    Micro Batch Size & Subset of the full batch for efficient gradient accumulation, influencing memory usage and computational efficiency. \\
    \midrule
    Fused Activation / Kernel Fusion & Integrates activation with preceding computation, reducing memory bandwidth usage and enhancing performance.\\
    \midrule
    Activation Recompute & Balances computation and memory by recomputing activations instead of storing them, affecting memory usage and computation time. \\
    \midrule
    Optimizer Offloading & Transfers optimizer states to CPU memory, conserving GPU memory, allowing for larger models but potentially slowing training. \\
    \midrule
    Weight Offloading & Stores model weights in CPU memory, transferring them to the GPU as needed, allowing for larger models but possibly impacting speed.\\
    \midrule
    Activation Offloading & Moves activations to CPU memory, saving GPU memory but potentially slowing training due to data transfer overhead.\\
    \midrule
    TP Compute/Comm Overlap & Overlaps tensor parallel computation and communication, improving efficiency but with varying implementation complexity.\\
    \midrule
    DP Compute/Comm Overlap & Overlaps data-parallel computation and communication, reducing communication overhead in data-parallel training.  \\
    \midrule
    (p2p)ReduceScatter \& Allgather / Allreduce & Communication patterns for synchronization and data exchange, affecting efficiency and scalability in distributed training. \\
    \midrule
    ZeRO-1 / Optimizer Sharding & Shards optimizer states among DP replicas, reducing memory usage. \\
    \midrule
    ZeRO-2 / Optimizer+Gradient Sharding & Shards optimizer states and gradients among DP replicas, further reducing memory usage. \\
    \bottomrule
    \end{tabular}%
}
\caption{Optimization Techniques and Their Impact on Training Performance}
\label{tab:optimization_techniques}
\end{table*}

\begin{figure}[htpb]

    \centering
    \includegraphics[width=0.4\textwidth]{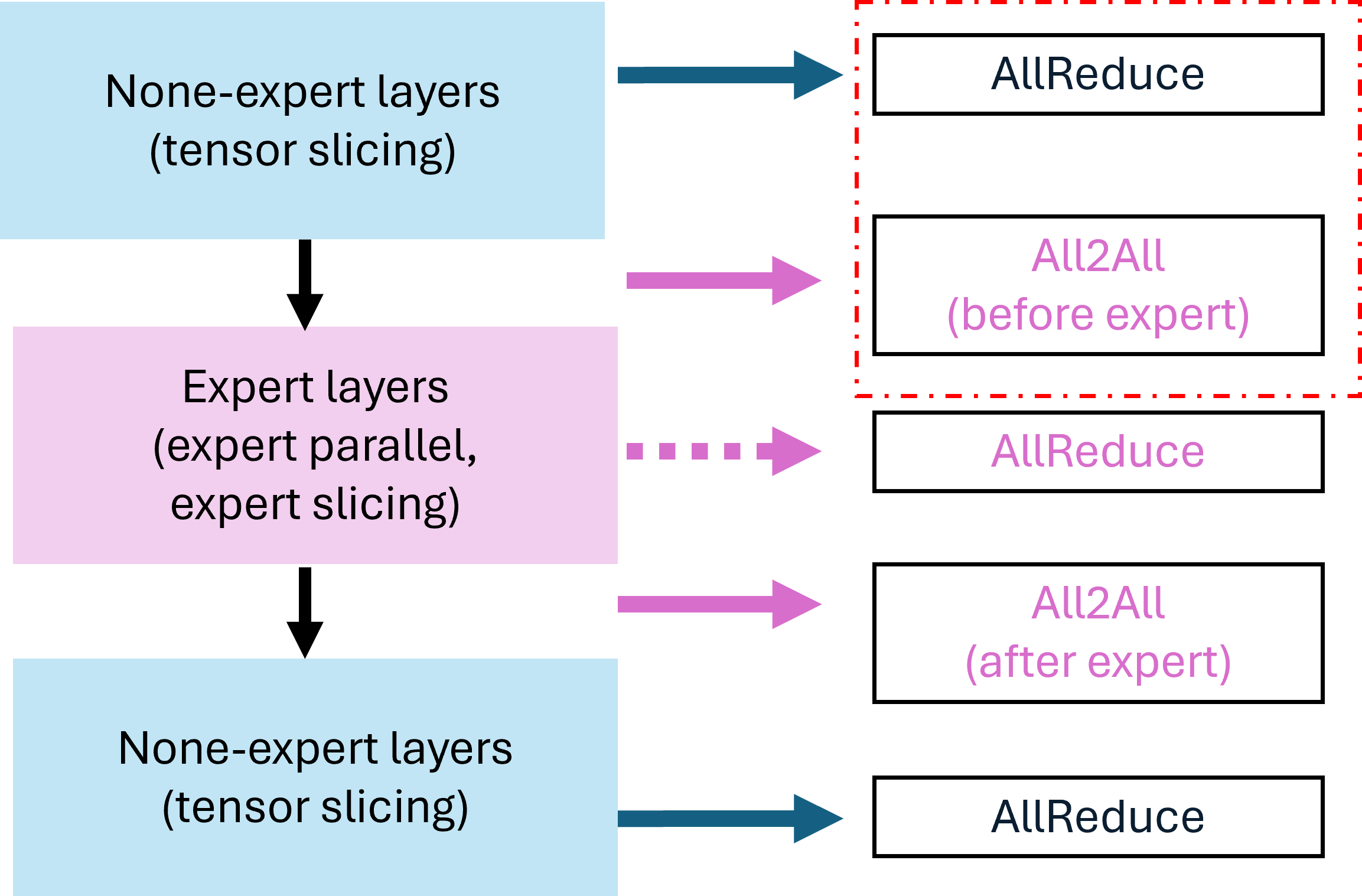} 
    \caption{MOE communication pattern.}
    \label{fig:moe-comm}
\end{figure}

Figure \ref{fig:moe-comm} illustrates MoE communication patterns. Expert parallelism requires extensive all-to-all communication, typically unnecessary for dense LLMs. Common communication and collective operations such as all-reduce, all-to-all, reduce-scatter, all-gather, and point-to-point communication can often be interleaved with computation with additional complexity.

Estimating optimal LLM performance is difficult due to the vast optimization space, which includes thousands to millions of parameter combinations from Table \ref{tab:optimization_techniques}. Empirical methods, like system testing, are often impractical for pinpointing the optimal configuration from all these combinations. As a result, programmers rely on intuition and published results, which can lead to suboptimal solutions. For instance, NEMO \cite{nvidia2025nemo}, a popular software framework uses a default setting of TP=ES and EP=\#Experts, deploying one expert per GPU. However, TP is limited by number of attention heads and feed-forward dimensions, while ES is not, making this strategy potentially suboptimal. This highlights the need for a comprehensive tool to predict the impact of optimization parameters and parallelization strategies. Our co-design study shows that a Fullflat network with same high bandwidth between any two nodes, high radix and low diameter, and low latency reduces performance gaps across different parameter combinations compared to conventional two-tiered networks (see Figure \ref{fig:opt}).

\subsection{Systems and Their Parameters}
In our exploration of future AI data center architectures, we focused on co-design alternatives involving two-tiered networks and full flat configurations. Two-tiered networks feature a low-latency, high-bandwidth domain (HBD) and a high-latency, low-bandwidth domain (LBD), also referred to as scale-up (SU) and scale-out (SO) domains, respectively \cite{DCT}, \cite{Cisco}. 

\begin{table}[hptb]
  \centering
  	\resizebox{0.4\textwidth}{!}{
    \begin{tabular}{lll}
     \toprule
    \textbf{Interconnect} & \textbf{BW/GPU} & \textbf{GPU Name} \\
	\midrule
    \textbf{NVLink 5.0} & 900 GB/s/dir & Blackwell \cite{wikipedia2025nvlink} \\
    \textbf{XeLink/Intel Scale-up} & 600 GB/s/dir & Gaudi3 \cite{tomshardware2024gaudi3} \\
    \textbf{Infinity Fabric} & 448 GB/s/dir & MI300X \cite{amd2024mi300x} \\
	\bottomrule
    \end{tabular}%
	}
	\caption{Existing scale-up interconnects.}
  \label{tab:scale-up-interconnect}%
\end{table}
The HBD consists of compute nodes or GPUs interconnected either directly or via a switch. For instance, Nvidia's H100 GPU connects eight GPUs through a NVSwitch, while the upcoming Rubin Vera will link 144 GPUs within a single HBD. Communication within an HBD utilizes high-bandwidth interconnects like NVLink, typically involving single-hop transfers, creating the illusion of unified shared memory across these GPUs. Table \ref{tab:scale-up-interconnect} lists current proprietary scale-up interconnects and their bandwidths. UALink \cite{grimm2024ualink} is an emerging initiative aimed at establishing an open interconnect standard for future systems. Communication between GPUs outside the HBD occurs through the LBD, requiring multiple hops and using lower bandwidth interconnects like standard Ethernet or UEC \cite{ultraethernet2025}, typically operating at 50 GB/s. We expect, by 2026, Ethernet speed could reach 100-200 GB/s, which we use as baseline for co-design study. 

\begin{figure}[tbph]
    \centering
	\includegraphics[width=0.15\textwidth]{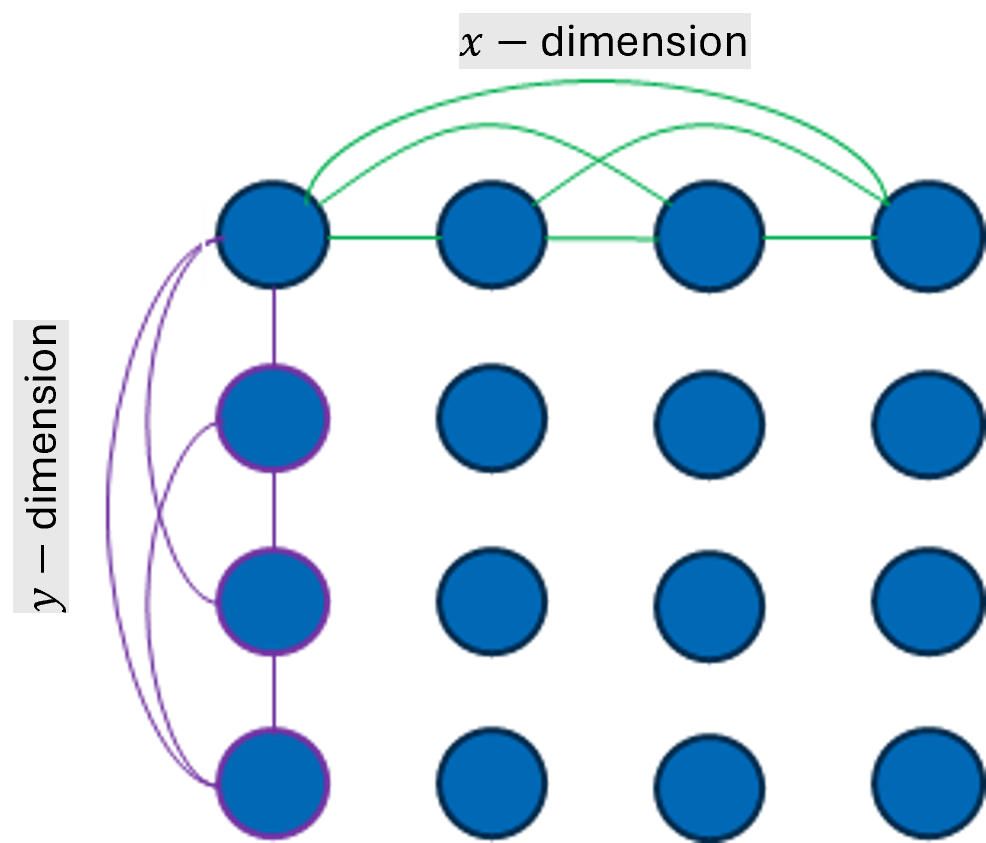}%
    \includegraphics[width=0.15\textwidth]{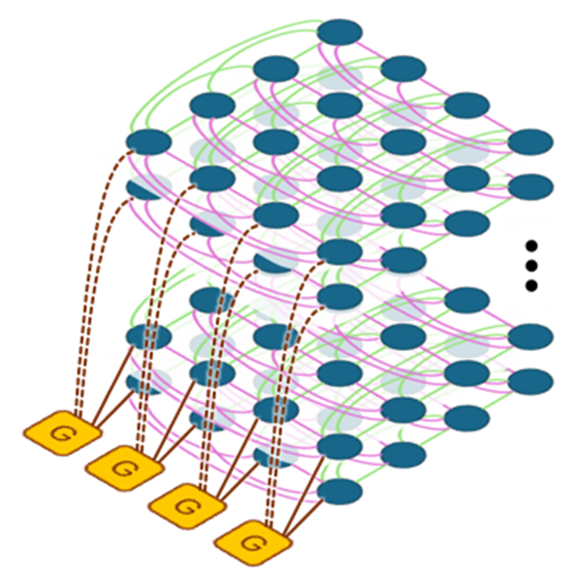}%
    \includegraphics[width=0.15\textwidth]{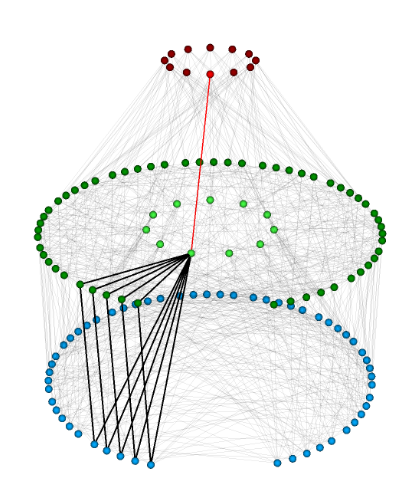}
    \caption{Network topologies: (left) 2D HyperX, (center) Rails of 2D HyperX switches with GPU (G) connected to each switch, (right) connected as Polarfly \cite{PolarFly-all}.}
    \label{fig:three-network-figures}
\end{figure}

Prior research \cite{Yin2013LIONS} and recent advancements \cite{Patel2025NVIDIA,Lightmatter2025PassageL200,Cerebras2025DARPA} suggest the feasibility of a fully optical networks with co-packaged optics (CPO) for both HBD/LBD domains, offering high bandwidth, reduced power consumption, and cost-effectiveness. These networks can adopt 2D or 3D configurations, such as HyperX or PolarFly, allowing access to any GPU in two or three hops. A Full flat network use advanced CPO to equalize bandwidth between HBD and LBD, though communication outside HBD incurs higher latency due to multiple hops. Figure \ref{fig:three-network-figures} shows a few potential Full flat topologies, with GPUs linked via single-hop connections within HBD, while those requiring multiple switches are part of LBD. FullFlat employs 2D/3D HyperX, Polar* or similar meshes, ensuring each node is \(\le\)3 hops from any other at equal high bandwidth per hop. Adaptive routing mitigates ``traffic storms" and reduces tail latency. FullFlat equalizes HBD and LBD (high and low bandwidth domains), allowing all traffic types on a single network without reconfiguration. This removes the scale-up/scale-out partitioning required in two-tier fabrics, simplifying scheduling and model mapping. Full flat refers to full bisection bandwidth over a folded high-radix, multi-stage optical mesh, not \(N^2\) physical links. FullFlat offers adaptive routing, flexible workload placement, and unique parallelization opportunities. It is less sensitive to missing optimizations (compute-comm overlap, hardware collectives), with only a 5\% performance gap across top 5000 configs (vs. 80\% for two-tier), which should translate to less performance engineering effort (\ref{fig:opt}). Additionally, it should also improve resiliency, robustness, and require less redundancy tax since any GPU can replace any faulty one anywhere in the network. While technical details of the Full flat network will likely be disclosed in a future paper, this study anticipates the development of such networks by 2026/2027.

With Nvidia envisioning a million-GPU AI factory supported by co-packaged optics (CPO) \cite{Moore2025CrucialOptical}, in this co-design study, we analyze a large data center housing approximately 65,536 GPUs, as we do not anticipate a single model needing larger system than this. We considered HBD sizes from 8 to 1024, reflecting current trends and future capabilities.

\begin{table}[hptb]
    \centering
	\resizebox{0.5\textwidth}{!}{
   \begin{tabular}{rrrr}
    \toprule
    \textbf{Architecture Specs} & \textbf{FullFlat} & \textbf{TwoTier-HBD 64|128} & \textbf{TwoTier-HBD8} \\
    \midrule
    \textbf{FLOPS FP8 (PF/s)} & 9.2   & 9.2   & 2 \\
    \textbf{FLOPS FP16 (PF/s)} & 4.6   & 4.6   & 1 \\
    \textbf{Tier 1 Memory BW (TB/s)} & 30    & 30    & 3 \\
    \textbf{Tier 2  Memory BW (GB/s)} & 256   & 256   & 450 \\
    \textbf{Tier 1 Memory Cap (GB)} & 432   & 432   & 80 \\
    \textbf{Tier 2 Memory Cap (GB)} & 480   & 480   & 512 \\
    \textbf{HBD Size} & 64|128    & 64|128    & 8 \\
    \textbf{Scale up BW (GB/s)} & 1600  & 1600  & 450 \\
    \textbf{Scale out BW (GB/s)} & 1600  & 200   & 50 \\
    \textbf{Tier 1 Latency (ns)} & 500   & 500   & 10000 \\
    \textbf{Tier 2 Latency (ns)} & 2000  & 2000  & 20000 \\
    \textbf{Cluster size} & 64K & 64K & 64K \\
    \bottomrule
    \end{tabular}%
	}
	    \caption{Architecture Specifications}
	\label{tab:sys-spec}
\end{table}

Table \ref{tab:sys-spec} outlines the system specifications used in our co-design analysis, informed by projections of future technological landscapes. The configuration {\emph{FullFlat}} refers to the full flat system with equal scale-up and scale-out bandwidths. {\emph{TwoTier-HBD64|128}} represents upcoming two-tiered network with around $8\times$ difference between scale-up and scale-out bandwidths, experimented with HBD sizes of 64 and 128. {\emph{TwoTier-HBD8}} reflects current systems with an HBD size of 8. The GPUs used for FullFlat and TwoTier-HBD64|128 are of same computational power and storage, whereas GPUs used in TwoTier-HBD8 are weaker. Our co-design study examines LLM performance sensitivity to various system parameters in Table \ref{tab:sys-spec}, providing insights into the system requirements for such models. We considered two-tiered networks with scale-up bandwidths of 450-3600 GB/s and scale-out bandwidths of 50-1600 GB/s, covering existing and potential future systems. The FLOPS FP8 and FP16 rows indicate floating-point operations per second for FP8 and FP16 precision, measured in petaflops. Tier 1 and Tier 2 Memory Bandwidth represent data transfer rates within the memory hierarchy, with Tier 1 denoting faster memory like HBM and Tier 2 being standard DDR/LPDDR memory. Memory Capacity reflects storage capacity in each tier. Scale-up and scale-out Bandwidth values describe intra-node/HBD and inter-node/LBD bandwidth capabilities. Latency indicates the minimum delay in data transfer for each network tier. Cluster Size denotes the number of nodes within each system. We anticipate systems of this scale will be available by 2026, and we analyze their performance on large-scale LLMs. 

\subsection{Tools for Performance Measurement}
This section describes a tool for performance projection that automates the exploration of a large parameter space in the early co-design stages. It evaluates system topologies and algorithmic optimizations to identify optimal combinations, aiding in hardware testing and complementing later detailed simulations.

To facilitate this, we extended an open-source tool Calculon \cite{isaev2023calculon} for our analysis. Calculon is an analytical tool designed for the high-level co-design of systems and LLM algorithms, providing insights into optimizations and the time, memory, and bandwidth required for optimal performance. It models parameters across:

\begin{itemize}[leftmargin=7pt]
\item {\emph{Application characteristics}}: Embedding size, sequence size, layer parameters, attention types, hidden layer sizes, feed-forward parameters, and batch size.
\item {\emph{Hardware system}}: Compute FLOPS, two-tier memory, and network configurations, with efficiency factors.
\item {\emph{Implementation strategies}}: Techniques like tensor parallelism (TP), sequence parallelism (SP), data parallelism (DP), pipeline parallelism (PP), batch size, micro-batch size, and strategies like 1F1B (one forward one backward) with interleaved pipeline stages. It includes activation/kernel fusion, recomputation, optimizer sharding, ZERO-1 and ZERO-2 optimizations, offloading weights, optimizers, activations, overlapping TP and DP communications, and choice of collective operations such as transitioning from all-reduce to reduce-scatter/all-gather, is considered\cite{ultrascale_playbook}.
\end{itemize}

Calculon analyzes diverse {\bf {dense}} LLM models, including Anthropic-52B, GPT-3 variants, Megatron, Turing, Chinchilla, Gopher, LaMDA, and Palm. Written in Python, Calculon develops a detailed analytical model for various LLM components, such as linear layers, batch matrix multiplication, ReLU, GeLU, dropout, softmax, batch normalization, TPComm (overlapped and non-overlapped), and element-wise operations. These models are then combined with the appropriate parameters into multi-head, multi-query attention blocks and MLP blocks. The complete LLM application is constructed by assembling layers of attention and MLP blocks as specified in a given model. This tool was validated on Megatron models \cite{shoeybi2019megatron} with 22B, 175B, 530B, and 1T parameters on the A100-based Selene supercomputer, achieving an error margin of ±10\% compared to actual performance.

We enhanced Calculon \cite{isaev2023calculon} to support {\bf{sparse}} Mixture of Expert (MoE) modeling for GPT-4 style models, incorporating expert parallelism (EP), expert sharding (ES), dynamic routing blocks, SwiGLU, select top-k block, expert communication, expert MLP blocks, and missing collective operations. The tool is based on real frameworks (e.g., Megatron + DeepSpeed), with configurable components for realistic MoE modeling. We integrated various hardware systems and LLM models, including GPT-4 and its variations, FullFlat, and TwoTier-HBD64|128, adapting TwoTier-HBD8 from the existing codebase. Scripts were added for performance projection and exhaustive search pipelines for MoE-based models. The Tool is modular; new model blocks (e.g., LORA\cite{bi2024deepseek}) and various MoE architectures can be added for further models for co-design studies. 

Our tool extension was validated using mistral \cite{mistral-models-overview} models on an 8-node cluster with 64 Nvidia H100 GPUs interconnected via NVLink 4.0 (HBD domain = 8), maintaining an error margin below ±10\%. Details on implementation, modeling, and validation will be discussed in a separate paper. For validation against NEMO \cite{nvidia2025nemo}, the tool was configured with one expert per GPU (EP=number of experts) and TP=EP=ES, reflecting NEMO's default setting, though this may not always optimize system utilization.

\section{Co-design Study}
This section presents our co-design studies, performance results, and insights. Table \ref{tab:moe_comparison} outlines the key parameters of the GPT models examined, including sparse MoE models GPT4-1.8T and GPT4-29T, and the dense model GPT3-175B. GPT-29T represents potential future large-scale models. For both MoE models, the topK value is set to 2, specifying the number of experts activated per token.

\begin{table}[tbph]
    \centering
	\resizebox{0.35\textwidth}{!}
	{
    \begin{tabular}{llll}
    \toprule
    \textbf{Model Name} & {\textbf{GPT-1.8T}}&{\textbf{GPT-29T}} & {\textbf{GPT3-175B}} \\
    \midrule
    {\textbf{\#Experts}} & 16    & 128   & 1 \\
    {\textbf{\#Layers}} & 120   & 120   & 96 \\
    {\textbf{Hidden Dim}} & 10752 & 15360 & 12288 \\
    {\textbf{FeedForward}} & 43008 & 61440 & 49152 \\
    {\textbf{\#Heads}} & 96    & 96    & 96 \\
    {\textbf{\#Key/Query}} & 112   & 160   & 128 \\
    \textbf{Model Params} & 1.8T  & 29T   & 175B \\
    \bottomrule
    \end{tabular}%
	}
    \caption{LLM Model Parameters. GPT 1.8T and GPT 29T are sparse MOE models, and GPT3-175B is a non-MOE dense model.}
    \label{tab:moe_comparison}
\end{table}

We modeled the performance of LLMs across various system configurations, using compute nodes that reflect both current and future systems. Employing an exhaustive search option, we identified optimal parameter configurations from Table \ref{tab:optimization_techniques} for the best runtime. We evaluated strong scaling on clusters of TwoTier-HBD* and FullFlat systems with 65,536 GPUs, assessing sensitivity to HBD size, network and memory bandwidth, capacity, and compute capabilities. Additionally, we explored the benefits of overlapping TP and DP communication with computation, and compared software versus hardware collectives. We analyze the impacts of system and application parameters on both sparse and dense models. The results provide insights into the advantages of Full flat optical networks and address the key question: {\emph{what type of data center is needed to support these models?}}

Inspired by FP8 training stability observed in DeepSeek-V3 \cite{bi2024deepseek}, we used FP8 as the primary data type in all experiments, while maintaining master weights and gradient accumulation in FP32. We assumed 80\% communication efficiency for both scale-up (Based on internal simulation and peer-reviewed Nvidia SHAAP results \cite{graham2020scalable}) and scale-out domains, 99\% flop efficiency for operations over size 128, and 90\% transfer efficiency from HBM for data sizes exceeding 100MB (benchmarked on Calculon). Efficiency decreases for smaller transfer sizes, as detailed in Calculon \cite{isaev2023calculon}. Table \ref{tab:notations} shows the notations used in the following.

\begin{table}[tbph]
  \centering
  \resizebox{0.4\textwidth}{!}
	{
    \begin{tabular}{lp{21em}}
	\toprule
    \textbf{Term} & \multicolumn{1}{l}{\textbf{Meaning}} \\
	\midrule
    HBD$x$  & There are $x$ compute engines in the HBD domain \\
    SU$n$   & $n$ GB/s Bandwidth for the scale-up/HBD domain \\
    SO$m$   & $m$ GB/s Bandwidth for the scale-out/LBD domain \\
	\bottomrule
    \end{tabular}%
	}
\caption{Notations and their meaning. BW means bandwidth.}
  \label{tab:notations}%
\end{table}%

\subsection{Strong Scaling with Cluster Size}

\begin{figure*}
    \centering
    \begin{subfigure}{0.24\textwidth}
        \includegraphics[width=\linewidth]{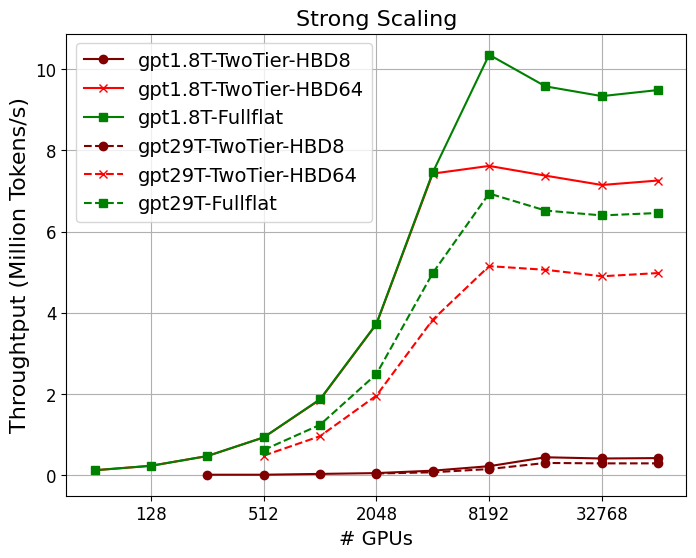}
        \caption{Strong Scaling}
        \label{fig:subfig1}
    \end{subfigure}\hspace{-5pt}
    \begin{subfigure}{0.24\textwidth}
        \includegraphics[width=\linewidth]{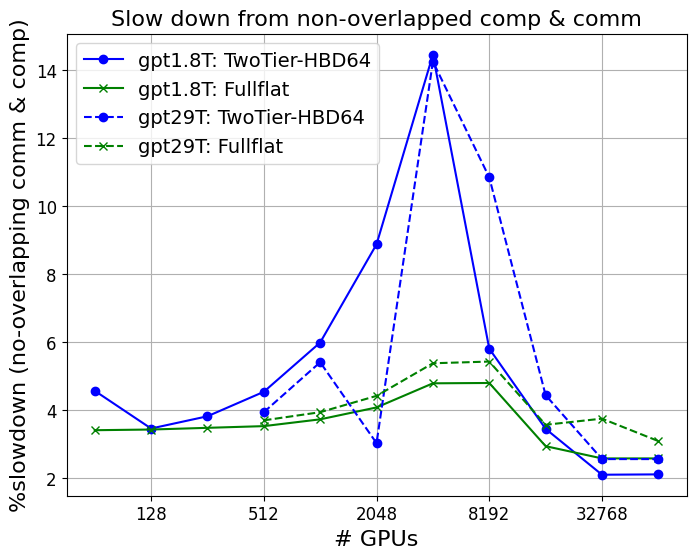}
        \caption{Non-Comm Overlap}
        \label{fig:subfig3}
    \end{subfigure}\hspace{-5pt}
    \begin{subfigure}{0.24\textwidth}
        \includegraphics[width=\linewidth]{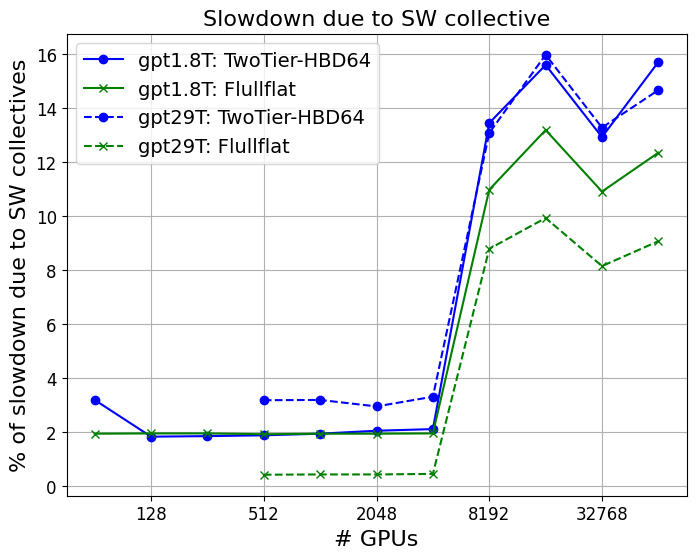}
        \caption{SW Collectives}
        \label{fig:subfig2}
    \end{subfigure}\hspace{-5pt}
    \begin{subfigure}{0.24\textwidth}
        \includegraphics[width=\linewidth]{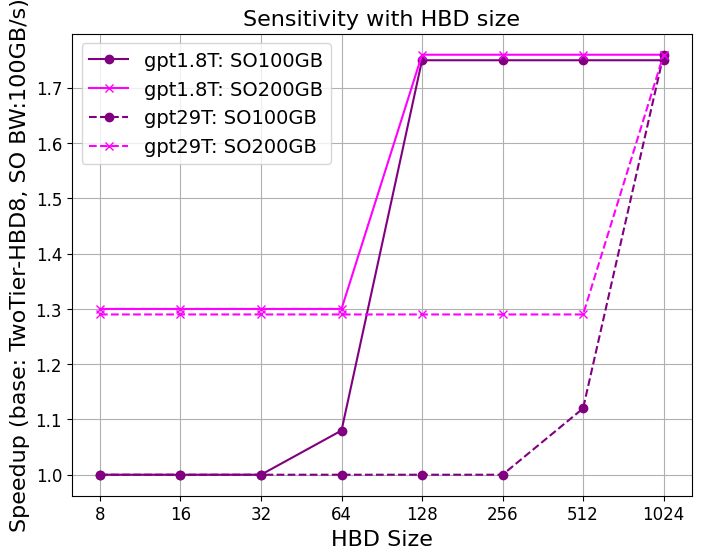}
        \caption{HBD Sensitivity (8192 GPUs)}
        \label{fig:subfig4}
    \end{subfigure}

    \vspace{-8pt}
    \begin{subfigure}{0.24\textwidth}
        \includegraphics[width=\linewidth]{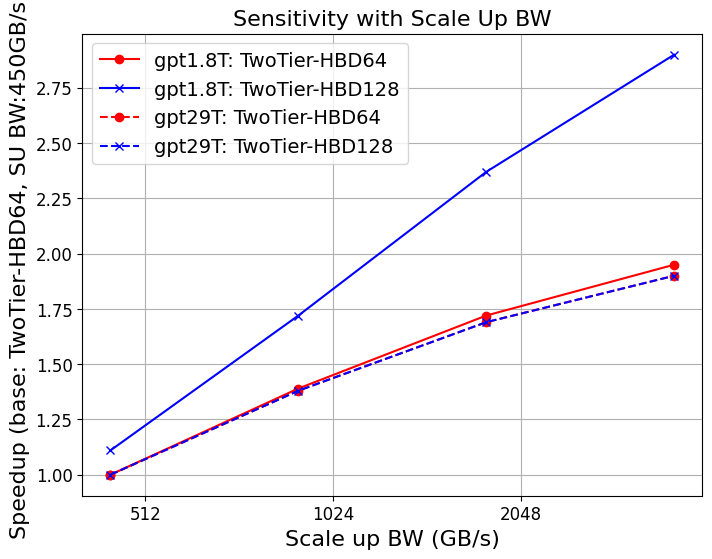}
        \caption{SU BW Sensitivity (8192 GPUs)}
        \label{fig:subfig7}
    \end{subfigure}\hspace{-5pt}
    \begin{subfigure}{0.24\textwidth}
        \includegraphics[width=\linewidth]{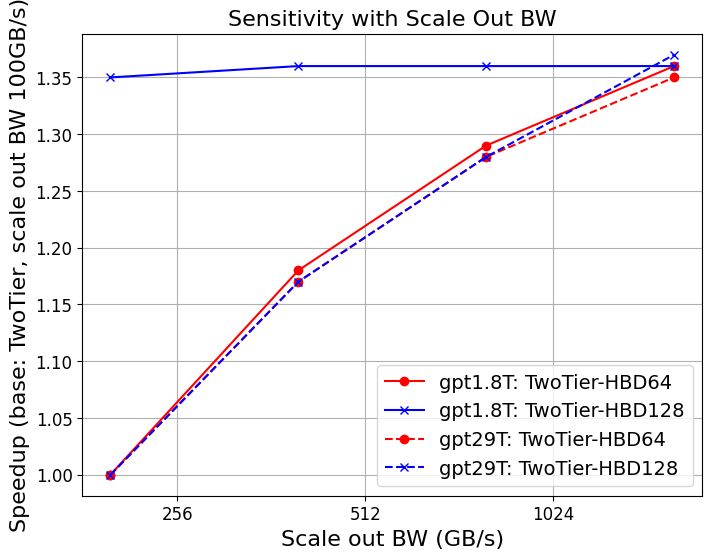}
        \caption{SO BW Sensitivity (8192 GPUs)}
        \label{fig:subfig8}
    \end{subfigure}
    \begin{subfigure}{0.24\textwidth}
        \includegraphics[width=\linewidth]{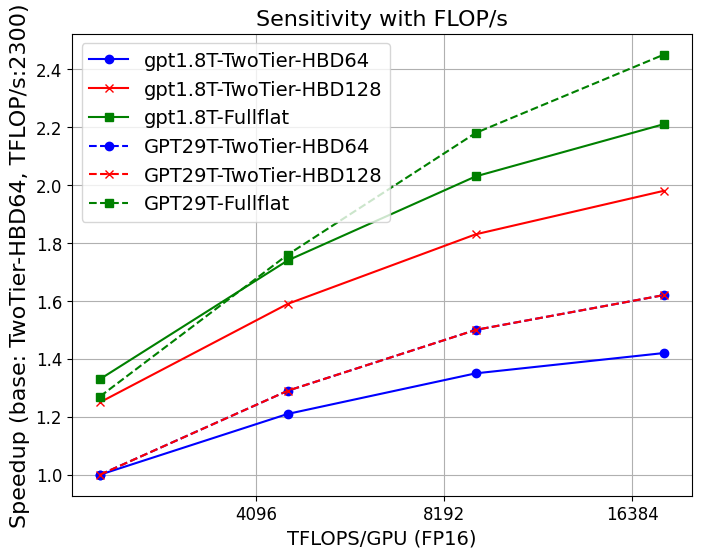}
        \caption{FLOP Sensitivity (8192 GPUs)}
        \label{fig:subfig6}
    \end{subfigure}\hspace{-5pt}
    \begin{subfigure}{0.24\textwidth}
        \includegraphics[width=\linewidth]{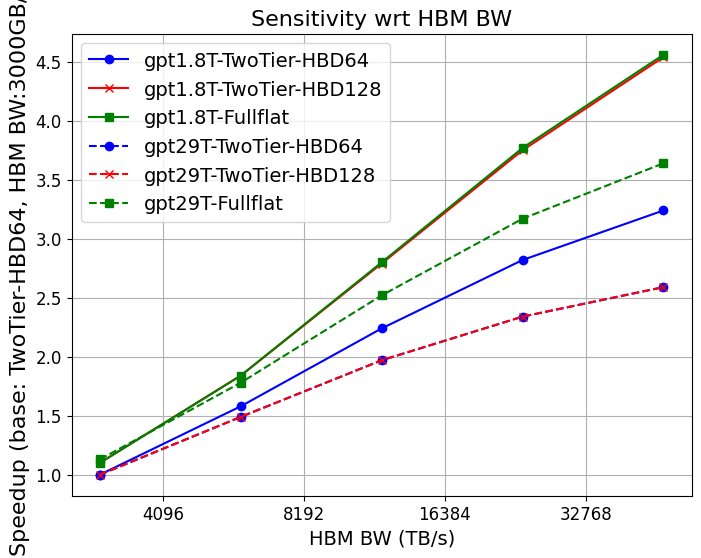}
        \caption{HBM BW Sensitivity (8192 GPUs)}
        \label{fig:subfig5}
    \end{subfigure}
	\hspace{-10pt}
    \caption{Performance trends for GPT4-1.8T (solid) and GPT4-29T (dashed) with sequence size 32,768 and batch size 1,024. Colors: Maroon=Two-TierHBD8, Blue=Two-TierHBD64, Red=Two-TierHBD128, Green=FullFlat.}
    \label{fig:gpt4}
	\hspace{-10pt}
\end{figure*}

Figure \ref{fig:gpt4} (a) shows throughput performance for GPT-4 MoE models with 1.8 trillion and 29 trillion parameters, measured in Million Tokens per second (MT/sec). Each batch contains 32,768 tokens, with a batch size of 1,024, totaling 32 million tokens. With a global batch size of 1024, DP can not scale beyond 1024. TP (and SP/ES) is limited by the number of attention heads and feed-forward dimensions. The x-axis represents the number of GPUs, ranging from 64 to 65,536, while the y-axis shows throughput. We compare three configurations:

\begin{itemize}[leftmargin=7pt] 
\item \textbf{TwoTier-HBD8 (current system):} With 80GB HBM per GPU, it requires 256 GPUs for GPT-1.8T and 2,048 GPUs for GPT-29T to accommodate the model parameters and activations. Throughput peaks at 16,384 GPUs before declining due to increased communication and other overheads. Increased parallelism (TP/SP/ES/PP) reduces compute time, leading to pipeline bubble overhead and exposed communication. For GPT-29T, the model size is so big that recomputation is always needed. More GPUs means higher parallelism but reduces per-GPU throughput. Tensor (TP) and Sequence (SP) Parallelism shard weights and activations and affect computation throughput of the LLM model. TP and TP+SP can't easily overlap with compute, making throughput reliant on communication bandwidth. Excessive TP can slow performance as can be observed in the plot \cite{ultrascale_playbook}. 

\item \textbf{TwoTier-HBD64 (near-future system):} Offers enhanced computational capabilities, larger HBD, increased memory/storage per GPU, and higher bandwidth. Throughput consistently increases, with minor degradation around 16,384 GPUs, plateauing up to 65,536 GPUs. Communication time starts dominating for GPT-29T (128 Experts) at 8K GPUs and  at 16K GPUs for GPT-1.8T (16 Experts). If the model size (and/or TP) exceeds HBD capacity, slower inter-node connectivity causes significant slowdowns. 

\item \textbf{FullFlat (future system):} This advanced CPO-based full flat system offers equal scale-up (HBD) and scale-out (LBD) bandwidths, greatly surpassing current capacities and delivering the highest throughput and better overall strong scaling. Its superior bandwidth minimizes communication dominance.

\end{itemize}

\begin{tcolorbox}[colframe=blue, colback=white, arc=1mm, boxrule=0.5pt, left=1mm, right=1mm, top=0mm, bottom=0mm, fontupper=\small, before skip=0.5mm, after skip=0.5mm]
{By 2026, systems are anticipated to run the GPT-1.8T model at speeds $50$-$70\times$ faster (at 4K gpus, TwoTier-HBD8 vs others), assuming advanced networking infrastructures are available. However, increasing the number of GPUs may lead to performance degradation beyond a certain point due to overheads compared to the pure compute being done. This underscores the importance of strategically investing in advanced network architectures for large-scale deployments of LLMs.}
\end{tcolorbox}

For GPT-1.8T, the performance of TwoTier-HBD64 and FullFlat is similar up to 4K GPUs, but diverges as communication shifts to scale-out (in order to utilize the added GPUs), resulting in a 30\% performance gap due to bandwidth differences. For GPT-29T, which consistently relies on scale-out for expert parallel communications, a 30\% performance gap exists between TwoTier-HBD64 and FullFlat, driven by scale-out bandwidth disparity and percentage of exposed communication in the application.

\subsection{Compute/Communication Overlap}

Figure \ref{fig:gpt4} (b) illustrates the slowdown from not overlapping compute and communication for TP and DP in GPT-1.8T and GPT-29T models on TwoTier-HBD64 and FullFlat systems. The y-axis shows percentage slowdown, while the x-axis shows the number of GPUs. For GPT-1.8T, TwoTier-HBD64 experiences a peak slowdown of 15\% at 4,096 GPUs, whereas FullFlat experiences 5\%, thanks to higher scale-out and flat HBD/LBD bandwidth, mitigating the performance loss. GPT-29T exhibits similar trends, with TwoTier-HBD64 peaking at 14\% and FullFlat at 4\%. The worst slowdown occurs with DP = 1024, TP = 4, and PP = 1, due to non-overlapped DP and TP communications. 

\begin{tcolorbox}[colframe=blue, colback=white, arc=1mm, boxrule=0.5pt, left=1mm, right=1mm, top=0mm, bottom=0mm, fontupper=\small, before skip=0.5mm, after skip=0.5mm] Overlapping compute and communication is important to reduce slowdowns as GPU count rises and associated overheads increase while computes per GPU decrease. Enhanced scale-up and scale-out bandwidth can help mitigate the negative effects of non-overlapping operations. A Fullflat network with equalized HBD/LBD bandwidth is less sensitive to this optimization compared to the two-tiered networks. \end{tcolorbox}

\subsection{Software vs. Hardware Collectives}
High-performance interconnects like NVIDIA Mellanox InfiniBand and Intel Tofino switches \cite{intel-tofino-tna} support hardware-accelerated in-network collectives for scale-out domains, while NVSwitches offer accelerated collectives for scale-up domains. Nvidia's SHARP FP8 technology allows CPUs and GPUs to offload collective tasks to the network chip \cite{Moore2025CrucialOptical,graham2020scalable} improving performance. The analytical model considers traffic volume, latency reduction, and GPU cycle savings (about 13\%) due to the presence of a hardware-supported collective engine. Figure \ref{fig:gpt4} (c) illustrates the performance slowdown from using software-based collectives, with $2\times$ traffic for all-reduce and $1.5\times$ for reduce-scatter and all-gather. The y-axis shows the percentage slowdown, with the number of GPUs varying from 64 (512 for GPT-29T) to 65,536 shown in the x-axis.

For both GPT4 MOE models, TwoTier-HBD64 experiences a peak slowdown of 16\% at 8,192 GPUs, highlighting the importance of hardware-accelerated collectives' at larger scales. FullFlat experiences a slowdown of 10\% to 13\%, indicating that higher equalized bandwidth in HBD/LBD mitigates some performance loss (by transferring extra traffic faster). The exhaustive search selects combinations that reduce the TP*DP factor to minimize collective costs in this setup.

\begin{tcolorbox}[colframe=blue, colback=white, arc=1mm, boxrule=0.5pt, left=1mm, right=1mm, top=0mm, bottom=0mm, fontupper=\small, before skip=0.5mm, after skip=0.5mm] Hardware collectives can improve performance by 16\% or more, with impacts more pronounced at higher GPU counts. FullFlat is less sensitive to missing optimizations of Hardware-based collectives. \end{tcolorbox}

\subsection{Sensitivity to {\bf{H}}igh {\bf{B}}andwidth {\bf{D}}omain}
Figure \ref{fig:gpt4} (d) shows the effect of different HBD sizes on GPT4 MOE models' performance using a two-tiered network with 8,192 GPUs. The x-axis represents HBD sizes (number of nodes connected with high bandwidth channels) from 8 to 1,024, while the y-axis indicates speedup relative to the smallest HBD size, with a 100 GB/s scale-out (SO) bandwidth. Full flat configuration has not been used in this experiment since for FullFlat, the entire cluster could be considered part of HBD and there is almost no sensitivity with HBD sizes.

Initially, there is a 30\% performance gap between the 100 GB/s and 200 GB/s SO configurations, attributed to bandwidth differences and percentage of exposed communication, which reduce communication time for SO200GB. Performance remains steady until HBD sizes are smaller than the product of expert parallelism (EP) and expert shard (ES). Significant improvements occur when expert parallelism can take full advantage of the higher bandwidth for all communications. Beyond this point, usually only data parallel (and/or PP) communications occur via the slower scale-out network, accounting for less than 1\% of the runtime.

\begin{tcolorbox}[colframe=blue, colback=white, arc=1mm, boxrule=0.5pt, left=1mm, right=1mm, top=0mm, bottom=0mm, fontupper=\small, before skip=0.5mm, after skip=0.5mm] 
Fast communication among experts in the MOE models is vital and would benefit from low-latency, high-bandwidth networks present in HBD domains. When a model's communication-intensive TP/EP/ES domains can fit in a smaller HBD environment, larger HBDs won't enhance performance. Data centers that can support fast communication among experts prevalent in common MoE LLM models will achieve better performance. Once expert communication fits within HBD, scale-out cost should drop sharply.
\end{tcolorbox}

The inflection points for SO100GB at HBD=64 for GPT-1.8T and HBD=512 for GPT-29T indicate that lowering tensor parallelism (TP), increasing pipeline parallelism (PP), and enabling attention-only recompute can be beneficial which minimizes TP communication overhead and improves runtime in certain cases. The upcoming Nvidia Rubin Ultra \cite{Patel2025NVIDIA} that comes with an HBD of 576 should be able to handle GPT-29T scale model well. 

\subsection{Sensitivity to Scale up (SU) Bandwidth}
Figure \ref{fig:gpt4} (e) shows the impact of scale-up (SU) bandwidth on GPT4 MOE model's performance, with scale-out (SO) bandwidth fixed at 200 GB/s. We examined HBD sizes 64 and 128. The x-axis represents SU bandwidth from 450 GB/s to 3,600 GB/s, while the y-axis shows the relative performance increase compared to the baseline: TwoTier-HBD64, SU bandwidth 450GB/s.

The most significant performance boost occurs when the SU bandwidth doubles from 450 GB/s to 900 GB/s. For GPT-1.8T with an HBD of 128, a $2\times$ increase in SU bandwidth results in a $1.6\times$ performance improvement. In contrast, with an HBD of 64, the improvement is $1.4\times$ because it relies on the slower SO network for some expert communications when using 8K GPUs. When the HBD is set to 128, experts can fully utilize the high-speed HBD/SU domain, leading to over a $2.62\times$ improvement from an $8\times$ increase in SU bandwidth. For GPT-29T, the performance is similar for both HBD64 and HBD128, as expert communication primarily depends on the SO domain in both scenarios. In this case, the $8\times$ increase in SU bandwidth provides a $1.9\times$ improvement.

\begin{tcolorbox}[colframe=blue, colback=white, arc=1mm, boxrule=0.5pt, left=1mm, right=1mm, top=0mm, bottom=0mm, fontupper=\small, before skip=0.5mm, after skip=0.5mm] 
Increasing SU bandwidth from 450 GB/s to 3,600 GB/s enhances performance across all HBD sizes, especially when expert and tensor parallel communications occur within the scale-up domain (HBD=128 for GPT-1.8T). However, diminishing returns suggest an optimal range for SU investments, beyond which resources may be better allocated to expanding the HBD domain and enhancing the scale-out network. 
\end{tcolorbox}

\subsection{Sensitivity to Scale-Out (SO) Bandwidth}
Figure \ref{fig:gpt4} (f) shows the impact of doubling scale-out (SO) bandwidth on GPT4 MOE models' performance, with scale-up bandwidth fixed at 1,600 GB/s at 80\% efficiency. We tested sensitivity for HBD sizes 64 and 128. For GPT-1.8T, in the HBD64 scenario, expert parallelization and communication occur through the scale-out domain, leading to consistent performance improvements with increased SO bandwidth ($1.36\times$ at 3.6TB/s). In the HBD128 case, since experts fit within HBD, improvements are limited to data and pipeline parallel communications with minimal impact (1\%) on overall performance.
For GPT-29T, in both HBD64 and HBD128 scenarios, most expert communications occur through the scale-out domain, resulting in steady performance gains with increased SO bandwidth ($1.35\times$ at 3.6TB/s). HBD128 shows slightly more improvements than HBD64 due to faster communication (i.e., all-to-all) within the HBD domain that the experts can leverage.
\begin{tcolorbox}[colframe=blue, colback=white, arc=1mm, boxrule=0.5pt, left=1mm, right=1mm, top=0mm, bottom=0mm, fontupper=\small, before skip=0.5mm, after skip=0.5mm] 
Increasing scale-out bandwidth significantly improves performance, especially in limited HBD configurations, highlighting its crucial role in expert communications. Optimizing HBD size is also essential to meet model requirements effectively. This conclusion should hold whether the communication across experts is balanced or imbalanced.
\end{tcolorbox}

\subsection{Sensitivity to Compute FLOPS}
Figure \ref{fig:gpt4} (g) illustrates the sensitivity of GPT4 MOE's performance to varying FLOPS per GPU. The SU bandwidth is fixed at 1.6 TB/s, and the SO bandwidth is set at 200 GB/s for the TwoTier-HBD* configuration. The relative speedup compared to the baseline: TwoTier-HBD64 at 2.3 FP16 TFLOP/s per GPU. As FLOPS increase from 2.3 PF/s to 18.4 PF/s, throughput improves across all configurations: TwoTier-HBD64, TwoTier-HBD128, and FullFlat. For the GPT-1.8T model, TwoTier-HBD64 sees a $1.4\times$ boost, while TwoTier-HBD128 and FullFlat improve by $1.59\times$ and $1.66\times$, respectively. The performance gap between TwoTier-HBD64 and TwoTier-HBD128 ranges from 25\% to 39\%. For the GPT-29T model, TwoTier-HBD64 and TwoTier-HBD128 exhibit performance gains around $1.62\times$, whereas FullFlat achieves a $1.9\times$ gain with increased FLOPS.

\begin{tcolorbox}[colframe=blue, colback=white, arc=1mm, boxrule=0.5pt, left=1mm, right=1mm, top=0mm, bottom=0mm, fontupper=\small, before skip=0.5mm, after skip=0.5mm] 
Increasing GPU FLOPS boosts performance, with larger HBDs and higher bandwidths offering the most significant gains, emphasizing the importance of HBD size and network bandwidth for optimizing LLM throughput. Thus, a balanced design of computational power and communication infrastructure is essential for optimal performance.\end{tcolorbox}

\subsection{Sensitivity to Memory Bandwidth}
Figure \ref{fig:gpt4} (h) shows the impact of varying HBM bandwidth from 3 TB/s to 48 TB/s across three configurations: TwoTier-HBD64, TwoTier-HBD128, and FullFlat, evaluated on 8,192 GPUs. For GPT-1.8T, increasing HBM bandwidth results in a $4.5\times$ performance improvement with a $16\times$ bandwidth increase. FullFlat consistently delivers the highest throughput, followed closely by TwoTier-HBD128, due to its capacity to accommodate most communications within the scale-up (HBD=128) domain for GPT-1.8T, resulting in less than a 1\% performance gap between TwoTier-HBD128 and FullFlat.

For GPT-29T, increasing HBM bandwidth enhances performance across all configurations, achieving up to $3.2\times$ improvement. FullFlat consistently gives the best performance, while TwoTier-HBD64 and TwoTier-HBD128 exhibit similar performance because both HBD domains cannot accommodate all TP/ES/EP communications in the SU domain, necessitating substantial use of the slower scale-out domain. 

\begin{tcolorbox}[colframe=blue, colback=white, arc=1mm, boxrule=0.5pt, left=1mm, right=1mm, top=0mm, bottom=0mm, fontupper=\small, before skip=0.5mm, after skip=0.5mm] This highlights the importance of HBM and network bandwidth in optimizing performance for LLMs. Larger HBD sizes and higher bandwidths yield significant throughput gains, emphasizing the need for balanced system configurations that integrate these elements effectively. 
\end{tcolorbox}

\subsection{Sensitivity to Memory Capacity}
\subsubsection{Impact of Tier-1 Memory:}
The main requirement for running an LLM is having sufficient memory in the system to store all necessary info. During training, various components must be stored in memory, including inputs, outputs, model weights, gradients, optimizer states, and activations. System kernels, like CUDA, may need additional 1 to 2 GB of GPU memory\cite{ultrascale_playbook}. Memory usage scales linearly with batch size and quadratically with sequence length, and models with MoEs require extra VRAM since all experts must reside in memory.

To evaluate the impact of per GPU HBM capacity on performance, we varied its size from 80 GB to 1,280 GB, including an infinite-size (large enough to fit a given model) scenario. Smaller memory sizes necessitate higher degrees of parallelism to partition the model to fit a single worker's (DP(PP(TP))) or EP(DP-Exp(PP(ES))) portion within a single GPU's memory. This added parallelism introduces communication and synchronization overheads and can reduce throughput. Having an infinite memory enables the use of only data parallelism in the attention layers and expert and data parallelism for MOE layers (if we choose to assign one GPU per expert as used as a default on several popular frameworks). Infinite memory also eliminates the need for recomputation (30\% cost reduction compared to full recompute) or model parameter offloading to CPU memory, reducing code complexity and associated performance overheads.

\begin{figure}[htbp]
    \centering
	\includegraphics[width=0.25\textwidth]{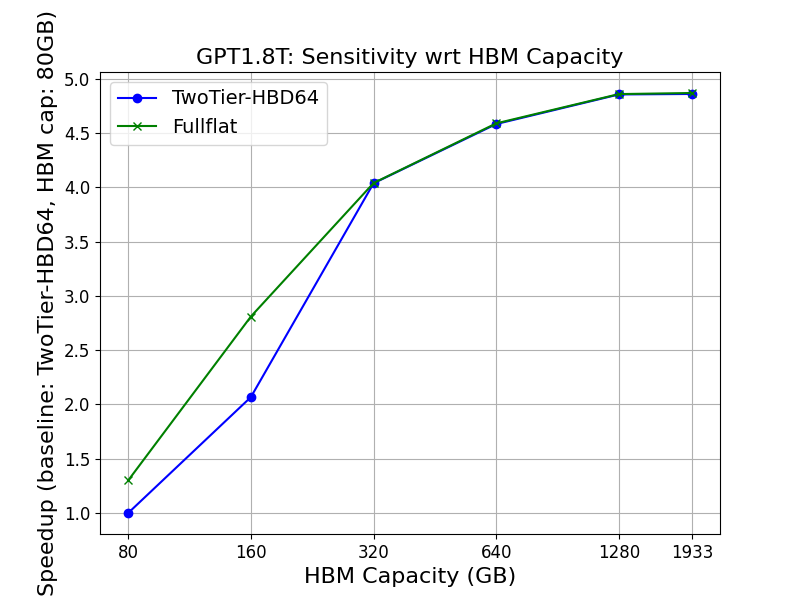}%
	\includegraphics[width=0.25\textwidth]{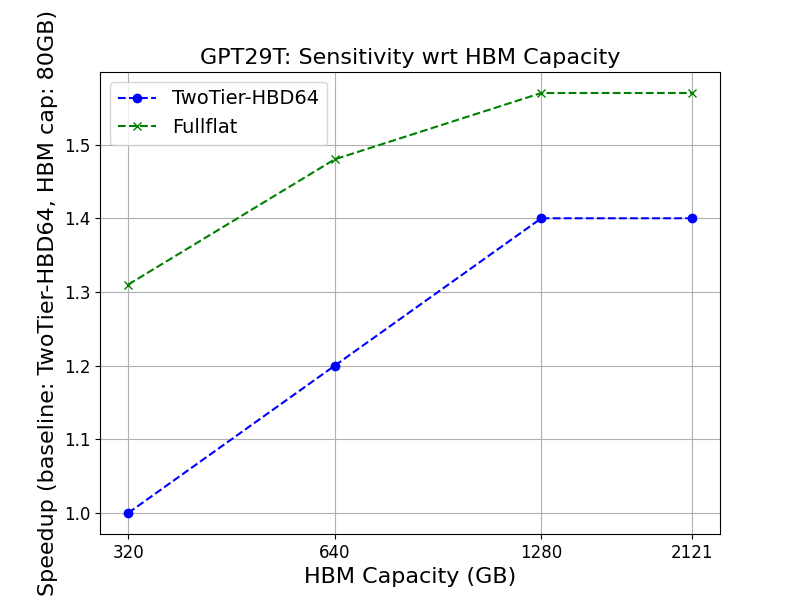}%
	\caption{Performance Impact of HBM Capacity. In this experiment, we used 512 GPUs, Seq size = 32,768, Batch size = 1024}
    \label{fig:mem_cap}
\end{figure}

Figure \ref{fig:mem_cap} shows the impact of HBM capacity on GPT4 models' throughput in TwoTier-HBD64 and FullFlat configurations using 512 GPUs. For GPT-1.8T, throughput improves as HBM capacity increases from 80 GB to infinite, with FullFlat consistently achieving the highest throughput. Gains are most pronounced up to 320 GB, after which they plateau, indicating limited benefits from additional HBM. At 1.28 TB, the model fits entirely, reducing parallelism to Data and Expert parallel domains. With 80 GB, a TP of 16 is used and SP/TP/EP/ES contributes 79\% to the total runtime. This reduces to 1.6\% as we increase the capacity to 640GB. The performance gap between TwoTier-HBD64 and FullFlat narrows from 30\% to 2\%, as both configurations share the same all-to-all bandwidth with 16 experts fitting in HBD64 for GPT-1.8T.

The GPT-29T model with 128 experts requires at least 320 GB of memory and a TP of 4. Expert and TP communication accounts for 37\% of runtime, and while higher HBM capacity improves throughput, gains are gradual due to high memory and communication needs. The FullFlat configuration excels, with performance gains stabilizing at 1.28 TB. In this setup, all-to-all communication is 12\% of runtime in TwoTier-HBD64, compared to 1.7\% in FullFlat. The performance gap between TwoTier-HBD64 and FullFlat is 30\% - 12\%, attributed to the scale-out domain's communication speeds: 200 GB/s for TwoTier-HBD64 and 1600 GB/s for FullFlat.

\begin{tcolorbox}[colframe=blue, colback=white, arc=1mm, boxrule=0.5pt, left=1mm, right=1mm, top=0mm, bottom=0mm, fontupper=\small, before skip=0.5mm, after skip=0.5mm] The benefit of co-design study is that it tells us exactly when to stop adding more resources in the system as it won't bring provide any more ROI. This analysis indicates that having a 1.3TB capacity is a decent target for data centers for the next couple of years. Increasing HBM capacity boosts throughput significantly ($4.9\times$ for GPT-1.8T). Thus, optimizing HBM capacity alongside other system parameters is crucial for achieving optimal performance for LLMs. \end{tcolorbox}

\begin{figure*}
    \centering	
	    \begin{subfigure}{0.24\textwidth}
        \includegraphics[width=\linewidth]{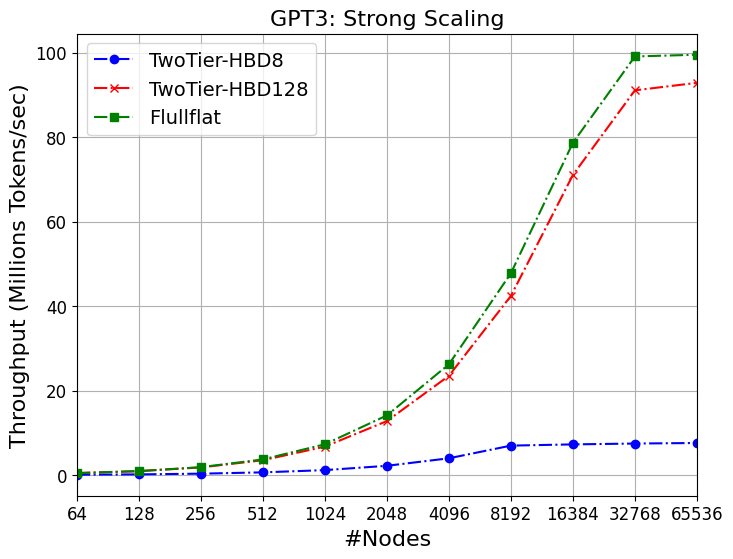}
        \caption{Strong Scaling}
        \label{fig:subfig1}
    \end{subfigure}\hspace{-5pt}
    \begin{subfigure}{0.24\textwidth}
        \includegraphics[width=\linewidth]{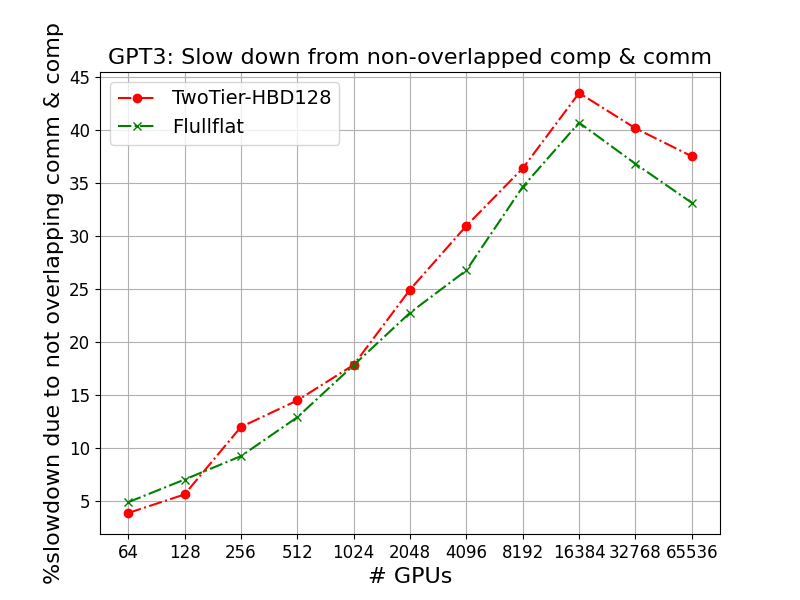}
        \caption{Non-Comm Overlap}
        \label{fig:subfig3}
    \end{subfigure}\hspace{-5pt}
    \begin{subfigure}{0.24\textwidth}
        \includegraphics[width=\linewidth]{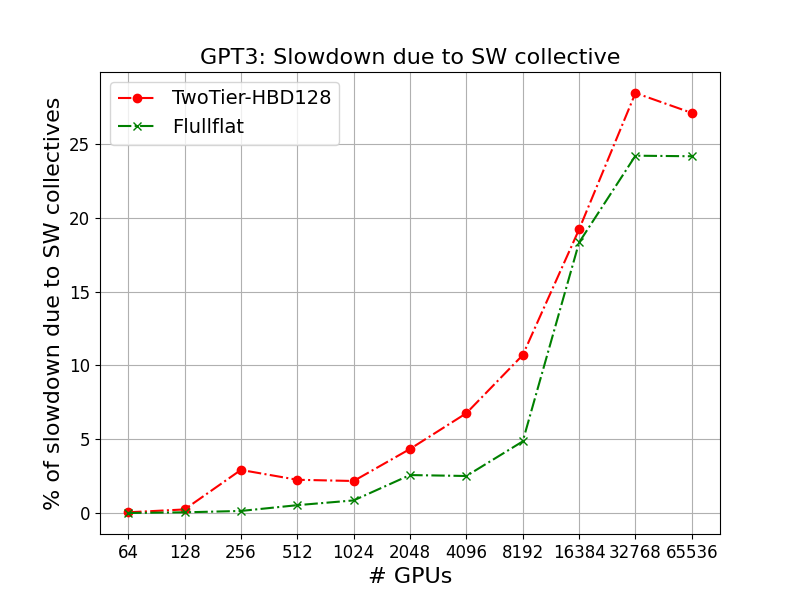}
        \caption{SW Collectives}
        \label{fig:subfig2}
    \end{subfigure}\hspace{-5pt}
    \begin{subfigure}{0.24\textwidth}
        \includegraphics[width=\linewidth]{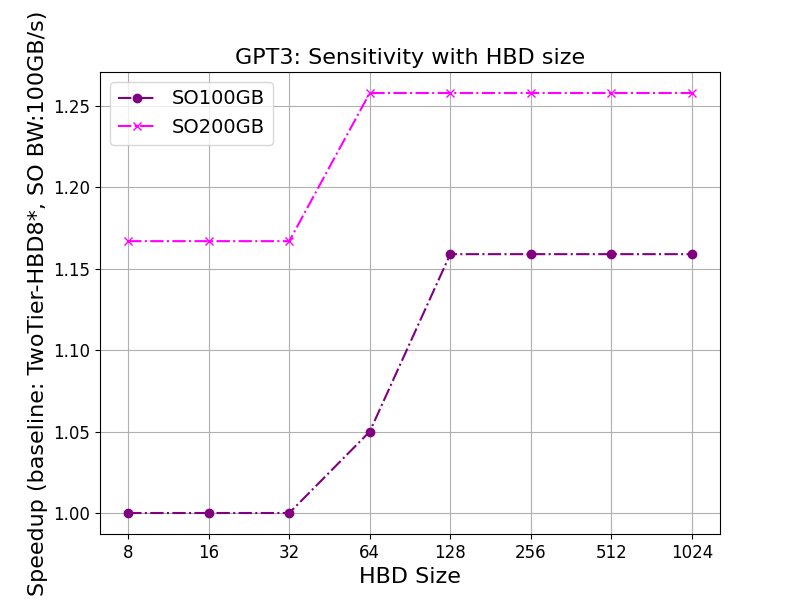}
        \caption{HBD Sensitivity (8192 GPUs)}
        \label{fig:subfig4}
    \end{subfigure}

    \begin{subfigure}{0.24\textwidth}
        \includegraphics[width=\linewidth]{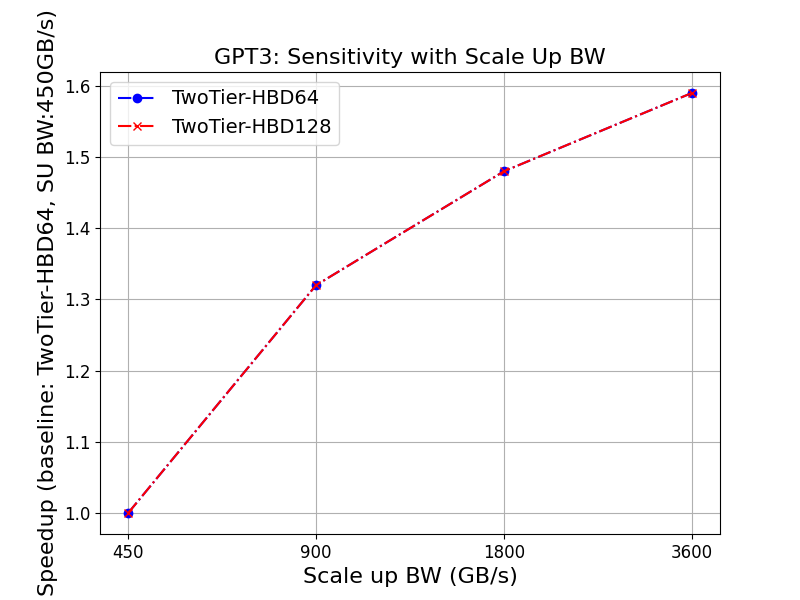}
        \caption{SU BW Sensitivity (8192 GPUs)}
        \label{fig:subfig7}
    \end{subfigure}\hspace{-5pt}
    \begin{subfigure}{0.24\textwidth}
        \includegraphics[width=\linewidth]{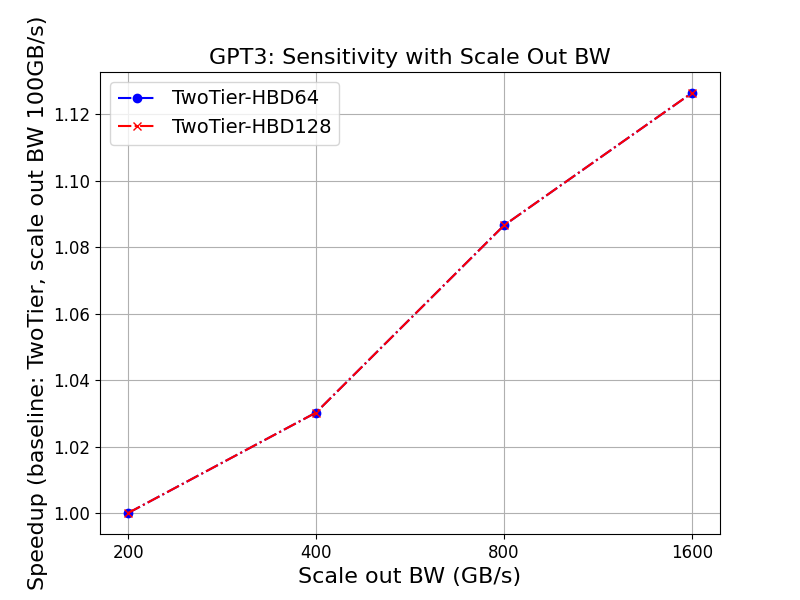}
        \caption{SO BW Sensitivity (8192 GPUs)}
        \label{fig:subfig8}
    \end{subfigure}
    \begin{subfigure}{0.24\textwidth}
        \includegraphics[width=\linewidth]{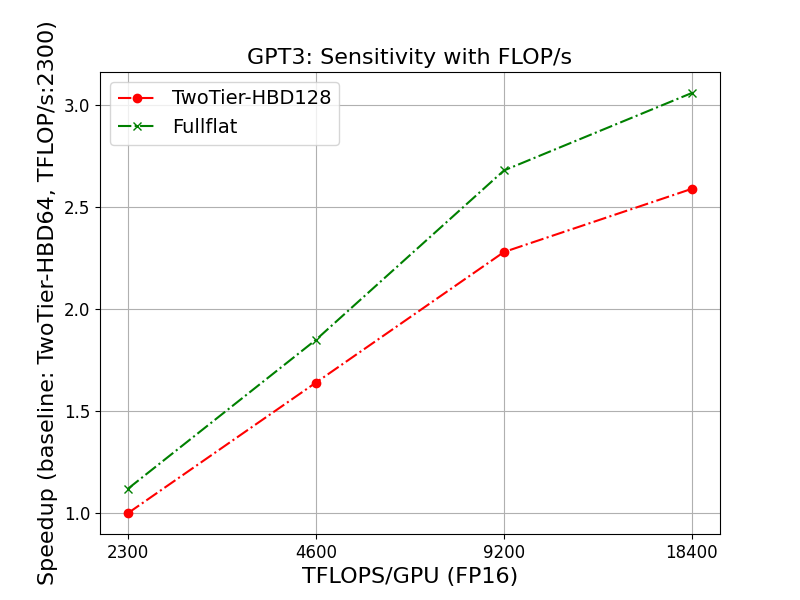}
        \caption{FLOP Sensitivity (8192 GPUs)}
        \label{fig:subfig6}
    \end{subfigure}\hspace{-5pt}
    \begin{subfigure}{0.24\textwidth}
        \includegraphics[width=\linewidth]{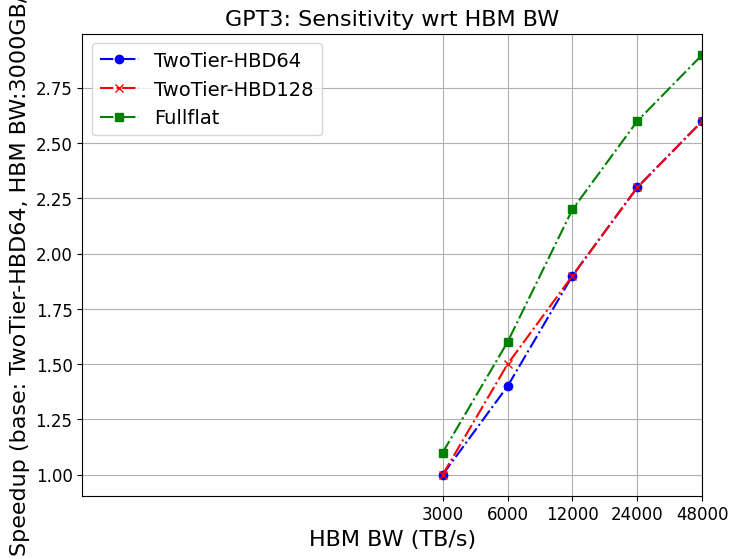}
        \caption{HBM BW Sensitivity (8192 GPUs)}
        \label{fig:subfig5}
    \end{subfigure}\hspace{-5pt}
    \caption{Performance trends for smaller, nonMOE, dense model GPT3-175B. Seq size = 2048, Batch size = 1024.}
    \label{fig:gpt3}
\end{figure*}

\subsection{Non MOE Dense LLM: GPT3-175B}
There is a recent trend towards distilling larger sparse MOE models into smaller dense models to achieve faster inference after deployment \cite{bi2024deepseek}. Thus, in co-design it's important to build a system that can serve both. To investigate how a data center, designed for sparse and trillion-parameter LLMs would perform with smaller dense non-MOE models, we conducted sensitivity experiments using the GPT-3 175 billion parameter model, which is significantly smaller than the GPT-4 MOE models. In general, dense models have higher arithmetic intensities, and thus, are less impacted by network bandwidths. However, one of the surprising findings of our analysis is that the denser GPT3 model is more sensitive to missing optimizations such as compute communication overlaps and missing hardware collectives.

Figure \ref{fig:gpt3} shows the GPT-3 175B model's performance sensitivity across different system setups. Throughput can be $10\times$ higher, with optimal scaling stabilizing at 32K GPUs. Performance is compute-bound until communication and other overheads become significant. Overlapping compute and communication is vital for smaller models, as neglecting this can cut performance by 43\% at 16,384 nodes. Lack of hardware collectives may lead to a 29\% drop. While HBD = 64 suffices for GPT-3, HBD = 128 is optimal for all models analyzed, given a scale-out bandwidth of 200 GB/s or more. This underscores the necessity for fast, efficient networks, hardware-accelerated collectives, and topology-aware communication libraries to effectively support LLM models in future data centers. 

\begin{table}[h!]
\centering
\resizebox{0.5\textwidth}{!}{
\begin{tabular}{|l|p{6cm}|p{6cm}|}
\hline
\textbf{Model} & \textbf{\%Exposed Comm Time} & \textbf{\%Overhead Time} \\
\hline
\textbf{GPT4-1.8T} &
\textbf{Average:} \textcolor{red}{TwoTier-HBD8 78\%}, \textcolor{blue}{TwoTier-HBD64 29\%}, \textcolor{deepgreen}{FullFlat 26\%} \newline
\textbf{Median:} \textcolor{red}{TwoTier-HBD8 91\%}, \textcolor{blue}{TwoTier-HBD64 17\%}, \textcolor{deepgreen}{FullFlat 17\%} &
\textbf{Average:} \textcolor{red}{TwoTier-HBD8 14.31\%}, \textcolor{blue}{TwoTier-HBD64 11.49\%}, \textcolor{deepgreen}{FullFlat 11.48\%} \newline
\textbf{Median:} \textcolor{red}{TwoTier-HBD8 0\%}, \textcolor{blue}{TwoTier-HBD64 0\%}, \textcolor{deepgreen}{FullFlat 0\%} \\
\hline
\textbf{GPT4-29T} &
\textbf{Average:} \textcolor{red}{TwoTier-HBD8 72\%}, \textcolor{blue}{TwoTier-HBD64 46\%}, \textcolor{deepgreen}{FullFlat 30\%} \newline
\textbf{Median:} \textcolor{red}{TwoTier-HBD8 91\%}, \textcolor{blue}{TwoTier-HBD64 42\%}, \textcolor{deepgreen}{FullFlat 22\%} &
\textbf{Average:} \textcolor{red}{TwoTier-HBD8 23\%}, \textcolor{blue}{TwoTier-HBD64 19\%}, \textcolor{deepgreen}{FullFlat 17\%} \newline
\textbf{Median:} \textcolor{red}{TwoTier-HBD8 2.6\%}, \textcolor{blue}{TwoTier-HBD64 0\%}, \textcolor{deepgreen}{FullFlat 0\%} \\
\hline
\textbf{GPT3-175B} &
\textbf{Average:} \textcolor{red}{TwoTier-HBD8 6.6\%}, \textcolor{blue}{TwoTier-HBD64 9.97\%}, \textcolor{deepgreen}{FullFlat 8.9\%} \newline
\textbf{Median:} \textcolor{red}{TwoTier-HBD8 6.5\%}, \textcolor{blue}{TwoTier-HBD64 3.92\%}, \textcolor{deepgreen}{FullFlat 2.29\%} &
\textbf{Average:} \textcolor{red}{TwoTier-HBD8 19\%}, \textcolor{blue}{TwoTier-HBD64 6\%}, \textcolor{deepgreen}{FullFlat 4\%} \newline
\textbf{Median:} \textcolor{red}{TwoTier-HBD8 2.98\%}, \textcolor{blue}{TwoTier-HBD64 2.86\%}, \textcolor{deepgreen}{FullFlat 0\%} \\
\hline
\end{tabular}
}
\caption{Percentage of exposed communication and overhead times. Full flat has the lowest percentage of exposed communication and overheads. This results also highlight that the dense non-MOE GPT3-175B has higher arithmetic intensity than sparse MOE GPT-1.8T and GPT-29T models.
}
\label{tab:comparison}
\end{table}

Table \ref{tab:comparison} presents the percentage of exposed communication and overhead times (recomputation, pipeline bubble, exposed offload times) for the three models discussed. MoE models exhibit significantly higher communication and overhead times compared to smaller dense models. The dense GPT3 model has higher arithmetic intensity and lower memory and network intensities compared to the MOE models. As a result, compared to the sparse MOE models, the dense model demonstrates a lower sensitivity to changes in scale-up and scale-out bandwidth. Sensitivity to HBM bandwidth increase is also lower (less than $3\times$ vs $4.5\times$ for the MOE model). However, flop sensitivity is higher than MOE counterparts ($3\times$ vs $2\times$). 

\begin{table}[t]
  \centering
  \resizebox{0.5\textwidth}{!}{
    \begin{tabular}{lp{7.785em}p{8.145em}p{7.855em}}
	\toprule
    & \multicolumn{3}{p{23.785em}}{\textbf{Impact Factor on FullFlat}} \\
	\toprule
    \textbf{System Component} & \multicolumn{1}{c}{\textbf{GPT4-1.8T}} & \multicolumn{1}{c}{\textbf{GPT4-29T}} & \multicolumn{1}{c}{\textbf{GPT3-175B}} \\
	\midrule
    \textbf{\#GPUs (128$\times$)} & 89$\times$ perf gain & 11$\times$ perf gain from 16$\times$ more GPUs & 100$\times$ perf gain  \\
    \textbf{No Comp-Comm Overlap} & 5\% slowdown & 4\% slowdown & 41\% slowdown \\
    \textbf{No HW-acclerated Collective} & 12\% slowdown & 10\% slowdown & 24\% slowdown \\
    \textbf{FLOPS (8$\times$)} & 1.66$\times$ perf gain & 1.93$\times$ perf gain & 2.73$\times$ perf  \\
    \textbf{HBM BW (16$\times$)} & 4.2$\times$ perf gain  & 3.24$\times$ perf gain & 2.63$\times$ perf gain \\
    \textbf{HBM Cap (2TB/GPU)} & 3.75$\times$ perf gain & 1.2$\times$ perf gain & N/A \\
	\toprule
    & \multicolumn{3}{p{23.785em}}{\textbf{Impact on TwoTier-HBD*}} \\
	\toprule
    \textbf{Scale UP BW (8$\times$)} & 2.62$\times$ perf gain & 1.9$\times$ perf gain & 1.59$\times$ perf gain  \\
    \textbf{Scale Out BW (8$\times$)} & 1.36$\times$ perf gain & 1.37$\times$ perf gain & 1.13$\times$ perf gain  \\
	\bottomrule
    \end{tabular}%
	}
  \caption{Impact Factor for Different System Components.}
  \label{tab:ImpactFactor}%
\end{table}

Table \ref{tab:ImpactFactor} summarizes the impact of system components on performance for FullFlat and TwoTiered systems, highlighting investment opportunities for data centers. The results encourage us to focus on Flops and scale-up bandwidth almost with equal importance, then HBM bandwidth, followed by compute and communication overlap, hardware-accelerated collectives, additional memory capacities, and scale-out bandwidth.

\subsection{Performance Benefit of FullFlat Network}

\begin{figure}
    \centering
    \includegraphics[width=0.4\textwidth]{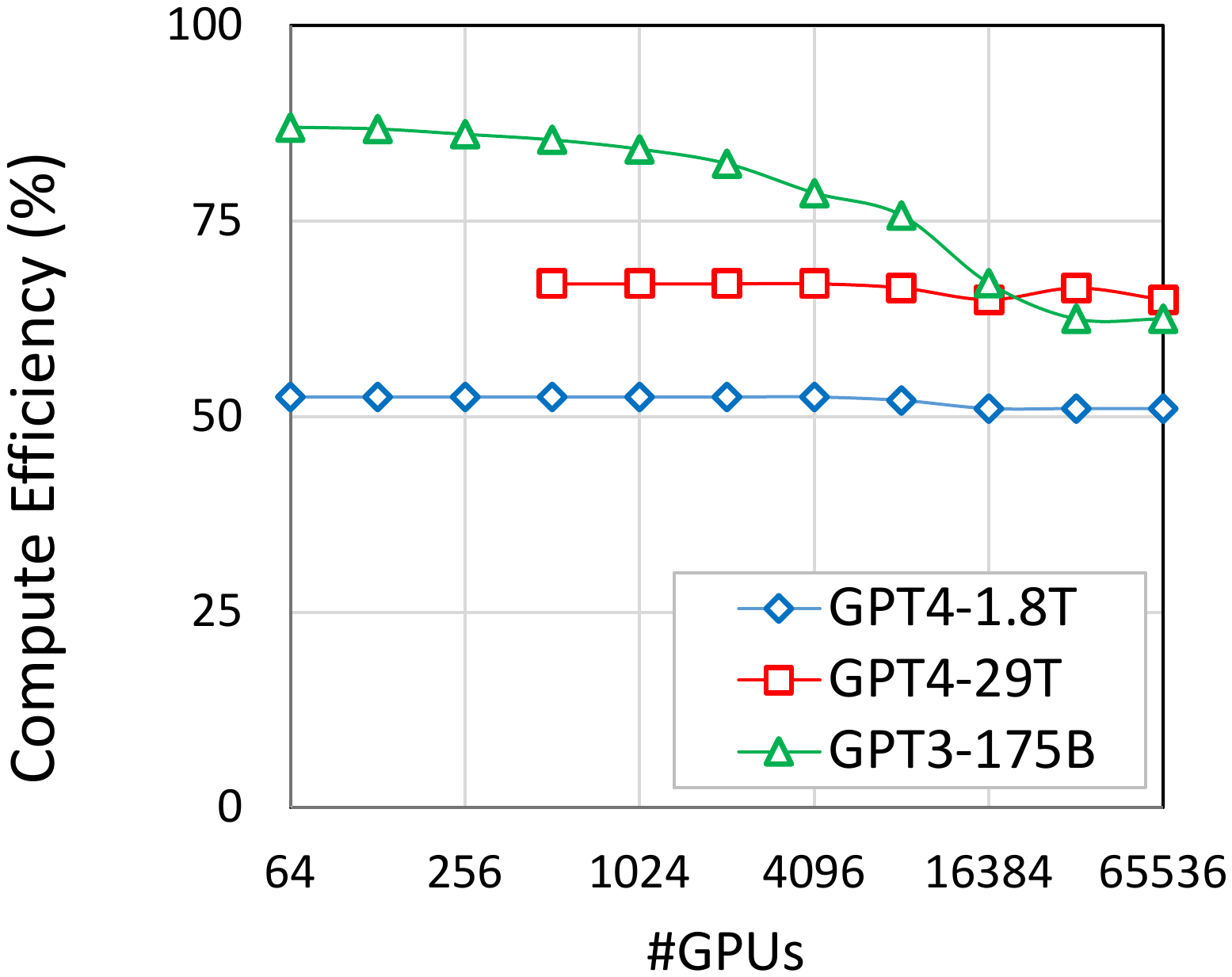} 
    \caption{Scaling of Compute Efficiency with Number of GPUs for FullFlat Configuration.}
    \label{fig:efficiency}
\end{figure}

Our studies show FullFlat networks offer superior performance and allows better utilization of increased resources. This setup allows unconventional strategies, such as using TP values going beyond a node boundary (suggested by popular wisdom), typically avoided in Two-Tier networks, especially with smaller HBD sizes. FullFlat with 1.6 TB/s bandwidth at 80\% efficiency improves performance by 30-60\% over Two-Tier networks with 200 GB/s scale-out bandwidth, enhancing MFU (see Figure \ref{fig:efficiency}) and system utilization to 70\% or more. Since FullFlat topology is less sensitive to missing optimizations, like comp/comm overlaps, users benefit from reduced programming effort, as the top 5,000 configurations show only a 5\% performance gap (see Figure \ref{fig:opt}). Programmers also need to worry less about achieving perfect comp/comm overlaps. 

A Full Flat network with advanced CPO offers improved reliability, availability, and serviceability. Fully optical interconnects are expected to experience fewer link flaps and would require fewer switches resulting in reduced power and costs \cite{Patel2025NVIDIA}. A Full Flat network improves resiliency, and robustness, and requires less redundancy tax since any GPU can replace any faulty one anywhere in the network. With equal HBD/LBD bandwidth and a high-radix, low-diameter, low-latency network, if a GPU fails in an HBD, it can be easily replaced with a spare from anywhere in the cluster, rather than replacing the entire HBD domain, improving TCO in the data center. While detailed TCO/power modeling is outside this paper’s scope, at a high level, CPO optics deliver \(\approx ~30-50\%\) lower GB/s than discrete optics, Co-packaged optics reduce per-port power by 30–50\% \cite{idtechex2024cpo,idtechex2024packaging,idtechex2024market}. Thus, FullFlat can reduce data center TCO and energy usage by up to 20\% or even more with innovations.

\section{Data Center We Need For LLM}
To effectively support LLMs, data centers must focus on high-performance computing, memory, and networking. Networks should aim for high-radix, low-latency interconnects with minimum bandwidths of 1.6 TB/s for scale-up and 200 GB/s for scale-out. Adopting full flat networks with integrated co-packaged optics and all-optical interconnects will reduce bottlenecks and improve the utilization of computational power and bandwidth.

The HBD should be large enough to support the number of experts in popular MoE models, targeting 20 FP16 PF/s per GPU (100 FP4 FP/s) for computational needs. Targeting 1.3 TB of HBM per GPU and 1-2 TB of Tier-2 memory capacity would be good for performance and would allow us to run very large models. It is recommended to have an HBM bandwidth of 30 TB/s or higher and a Tier-2 memory bandwidth of 256 GB/s. Integrating hardware-accelerated collective operations can enhance performance by minimizing communication volume. 

Data centers should include software packages that support established optimization strategies. They should support quick static Ridgeline \cite{Ridgeline} analysis to select optimal configurations for efficient resource utilization, maximizing MFU, and ensuring effective compute-communication overlaps while satisfying all other constraints. 

\section{Conclusion}
This study emphasizes the importance of co-designing data center architecture to effectively support large-scale large language models (LLMs). It highlights how these designs can uncover unexpected breakpoints, such as scenarios where adding a specific resource does not yield a higher return on investment (ROI). Our findings demonstrate the advantages of full flat network designs, which help reduce communication bottlenecks, improve system utilization, and decrease performance sensitivity to missing optimizations. This makes performance optimization less burdensome.

Balancing computational power, memory capacity, and network bandwidth is essential for achieving optimal performance. Therefore, future data centers should prioritize high-radix, low-latency interconnects and ample memory resources to accommodate the requirements of trillion-parameter models. Additionally, implementing hardware-accelerated collectives and strategic parallelism can help minimize slowdowns and maximize Model FLOPS Utilization (MFU).

It is also crucial to optimize software stacks to select the best parameter configurations for optimal compute and communication overlaps. These insights provide a roadmap for developing next-generation data centers that can efficiently deploy complex and resource-intensive models, ensuring ongoing advancements in AI capabilities.

\paragraph{Acknowledgement:}
\scriptsize{
We acknowledge the use of GPT technology and grammarly for enhancing the clarity and coherence of our writing.} 
\newpage
\bibliographystyle{ACM-Reference-Format}
\bibliography{sample-base}

\section{Appendices}

\subsection{Optimal Parameters}
\begin{table*}[htbp]
  \centering
  \resizebox{\textwidth}{!}{
    \begin{tabular}{cccccccccccccc}
    \toprule
    \textbf{Config} & {\textbf{TP}} & {\textbf{PP}} & {\textbf{DP}} &{\textbf{\#Experts}} & {\textbf{K}} & {\textbf{EP}} & {\textbf{ES}} & {\textbf{DP\_exp}} & \textbf{TP comm type} & \textbf{TP overlap} & {\textbf{DP overlap}} & \textbf{Recomp.} & {\textbf{ZERO-2}} \\
    \midrule
    \textbf{TwoTier-HBD8} & 16    & 1     & 256   & 16    & 2     & 16    & 16    & 16    & ar    & ring  & TRUE  & None  & TRUE \\
    \textbf{TwoTier-HBD64} & 4     & 1     & 1024  & 16    & 2     & 16    & 4     & 64    & ar    & ring  & TRUE  & None  & TRUE \\
    \textbf{Fullflat} & 4     & 1     & 1024  & 16    & 2     & 16    & 4     & 64    & ar    & ring  & TRUE  & None  & TRUE \\
    \bottomrule
    \end{tabular}%
	}
	  \caption{Optimal Parameters Picked By The Tool at 4K Nodes for GPT4-1.8T.}
  \label{tab:gpt1.8T-opt-param-4k}%
\end{table*}%

\begin{table*}[htbp]
  \centering
  \resizebox{\textwidth}{!}{
    \begin{tabular}{cccccccccccccc}
    \toprule
    \textbf{Config} & {\textbf{TP}} & {\textbf{PP}} & {\textbf{DP}} & {\textbf{\#Experts}} & {\textbf{K}} & {\textbf{EP}} & {\textbf{ES}} & {\textbf{DP\_exp}} & \textbf{TP comm type} & \textbf{TP overlap} & {\textbf{DP overlap}} & \textbf{Recomp.} & {\textbf{ZERO-2}} \\
    \midrule
    \textbf{TwoTier-HBD8} & 16    & 1     & 512   & 128   & 2     & 128   & 16    & 4     & ar    & ring  & FALSE & attn\_oly & TRUE \\
    \textbf{TwoTier-HBD64} & 8     & 1     & 1024  & 128   & 2     & 128   & 8     & 8     & ar    & ring  & TRUE  & None  & TRUE \\
    \textbf{Fullflat} & 8     & 1     & 1024  & 128   & 2     & 128   & 8     & 8     & ar    & ring  & TRUE  & None  & TRUE \\
    \bottomrule
    \end{tabular}%
	}
	\caption{Optimal Parameters Picked By The Tool at 8K Nodes for GPT4-29T.}
  \label{tab:gpt29T-opt-param-8k}%
\end{table*}%

\begin{table*}[htbp]
  \centering
   \resizebox{\textwidth}{!}{
    \begin{tabular}{ccccccccccccc}
    \toprule
    \textbf{Config} & {\textbf{TP}} & {\textbf{PP}} & {\textbf{DP}} & {\textbf{PP interleave}} & \textbf{TP comm type} & \textbf{TP overlap} & {\textbf{DP overlap}} & \textbf{Recomp.} & {\textbf{ZERO-2}} & {\textbf{Weight offload}} & {\textbf{Activations offload}} & {\textbf{Optimizer offload}} \\
    \midrule
    \textbf{TwoTier-HBD8} & 8     & 8     & 256   & 12    & rs\_ag & ring  & TRUE  & None  & TRUE  & TRUE  & TRUE  & TRUE \\
    \textbf{TwoTier-HBD64} & 16    & 8     & 128   & 12    & rs\_ag & ring  & TRUE  & None  & FALSE & FALSE & FALSE & FALSE \\
    \textbf{Fullflat} & 16    & 2     & 512   & 48    & rs\_ag & ring  & TRUE  & None  & TRUE  & FALSE & FALSE & TRUE \\
    \bottomrule
    \end{tabular}%
	}
	  \caption{Optimal Parameters Picked By The Tool at 16K Nodes for GPT3-175B.}
    \label{tab:gpt3-opt-param-16k}%
\end{table*}%

\end{document}